\documentclass[nofootinbib]{revtex4}

\usepackage{amsfonts, amsmath, graphicx, psfrag, array, bbm, float}

\newcommand{\be}{\begin{equation}}
\newcommand{\ee}{\end{equation}}

\newcommand{\bes}{\begin{eqnarray}}
\newcommand{\ees}{\end{eqnarray}}

\newcommand{\ba}{\begin{array}}
\newcommand{\ea}{\end{array}}

\newcommand{\so}{\mathfrak{so}}
\newcommand{\su}{\mathfrak{su}}
\newcommand{\SO}{\textrm{SO}}

\newcommand{\cM}{{\cal M}}
\newcommand{\cP}{{\cal P}}
\newcommand{\cS}{{\cal S}}

\newcommand{\di}{{}_i}
\renewcommand{\dj}{{}_j}

\newcommand{\us}{{}^1\!}%{{}^{\mathcal{S}}\!}
\newcommand{\uo}{{}^2\!}%{{}^{\mathcal{O}_1}\!}
\newcommand{\uoe}{{}^3\!}%{{}^{\mathcal{O}_2}\!}

\begin{document}

%%%%%%%%%%%%%%
% Title page
%%%%%%%%%%%%%%

\title{\bf Simplicity in simplicial phase space}

\author{Bianca Dittrich}\email{dittrich AT aei.mpg.de}
\author{James P. Ryan}\email{james.ryan AT aei.mpg.de}
\affiliation{ {\footnotesize MPI f\"ur Gravitationsphysik, Albert Einstein Institute,  Am M\"uhlenberg 1, D-14476 Potsdam, Germany}}

\begin{abstract}
A key point in the spin foam approach to quantum gravity is the implementation of simplicity constraints in the partition functions of the models. Here, we discuss the imposition of these constraints in a phase space setting corresponding to simplicial geometries. On the one hand, this could serve as a starting point for a derivation of spin foam models by canonical quantisation. On the other, it elucidates the interpretation of the boundary Hilbert space that arises in spin foam models.

More precisely,  we discuss different versions of the simplicity constraints, namely gauge-variant and gauge-invariant versions. In the gauge-variant version, the primary and secondary simplicity constraints take a similar form to the reality conditions known already in the context of (complex) Ashtekar variables. Subsequently, we describe the effect of these primary and secondary simplicity constraints on gauge-invariant variables.   This allows us to illustrate their equivalence to the so-called diagonal, cross and edge simplicity constraints, which are the gauge-invariant versions of the simplicity constraints. In particular, we clarify how the so-called gluing conditions arise from the secondary simplicity constraints.  Finally, we discuss the significance of degenerate configurations, and the ramifications of our work in a broader setting.

\end{abstract}

\maketitle

%%%%%%%%%%%%%%
% Introduction
%%%%%%%%%%%%%%

\section{Introduction}

In the quest to find a theory of quantum gravity, a class of theories arise, whose respective Hilbert spaces are labelled by graphs.  This class includes loop quantum gravity \cite{lqg}, spin foams \cite{sf}, quantum Regge calculus \cite{qRc}, causal dynamical triangulations \cite{cdt} and so on.  In loop quantum gravity, which is a canonical quantisation of continuum general relativity, these graphs (or spin networks as they are known) arise naturally and provide a basis for the space of states prior to the imposition of the Hamiltonian constraint. 
 On the contrary, in spin foams and quantum Regge calculus, the graphs take on a  greater significance, implementing a discretisation of the continuum data.  In these quantisation methods, one views the abstract graph as dual to a simplicial manifold, upon which the gravitational theory is expressed.\footnote{For an alternative view on spin foams, in particular allowing for graphs not dual to a simplicial manifold, see \cite{lkk,bahr}.} Thus, it acts as a uv--regulator which naturally cuts out certain divergences from the quantisation.  There is an obvious trade-off here.  Altering the fundamental component of space-time, the manifold itself, in a theory meant to describe the quantum dynamics of that very structure means that one must tackle anew some subtle technical issues.   

Already at the classical level, one needs to formulate a consistent lattice dynamics, and furthermore,  one should understand the significance and extent to which the lattice structure affects a realisation of the symmetries of the theory \cite{gambinipullin,diffeodiscrete}.  Even before dealing with this issue, upon a simplicial manifold, the ease with which one can formulate an action principle and endow the topological structure with gravitational data depends heavily upon how one parameterises the theory.  A natural way to encode a simplicial metric geometry is to use edge-lengths.  Indeed, this is the underlying structure of Regge's original proposal.  These are not the variables one uses, however, in the framework upon which modern spin foam quantisation rests.   

Spin foam quantisation is implemented on a class of theories that emerge as the lattice counterparts of non-metric formulations of general relativity, in particular the Plebanski formulation \cite{plebsf}.  In the continuum, these actions are functionals of a bi-vector, a spin connection and Lagrange multipliers that ensure the solution space includes that of the metric theory.\footnote{Equivalence between metric and non-metric formulations can typically only be obtained if certain non--degeneracy conditions hold. As we shall see these conditions will be particularly important in a discretisation.}  These actions take the form of constrained topological field theories, where the unconstrained action describes $BF$ theory.  The advantage is that one circumvents initial stumbling blocks which plague other formulations.  The path integral quantisation of $BF$ theory on a simplicial manifold is well understood.  The measure on the space of paths is well-defined and anomaly-free while the weight for each path is given succinctly by a lattice form of the action. Since any quantum amplitude is a topological invariant, one may view it as a well-defined prescription for the amplitude on {\it any} simplicial or continuum manifold of the same topology.  The aim is to devise a similar strategy for the constrained theory.  First one must devise a system of classical constraints for a simplicial manifold which restrict the topological theory to the gravitational theory.  Then, one must impose those constraints consistently at the quantum level.  This will alter both the measure and the weights of the original topological amplitude.  

The process is non-trivial for a multitude of reasons and constitutes the key point in the construction of spin foam models. Different methods to impose the constraints in the discrete setting have been developed and in fact, this partially constitutes the differences among these existing models \cite{barrett-crane,epr,fk,livs,daniele,valentin}. In a canonical analysis of the continuum Plebanski action, both primary and secondary simplicity constraints arise.  The secondary set ensure that the primary set are preserved under time-evolution.   Most of the aforementioned methods implement only a discrete version of the primary simplicity constraints explicitly in the path integral amplitudes. In contrast, \cite{sergei} argues that one should impose both the primary and secondary simplicity constraints, as this would ensure the correct correlation functions for all observables.  This proposal has not been brought to completion yet, as a suitable discretisation for the secondary simplicity constraints was missing \cite{henneaux}. Additionally, the secondary simplicity constraints involve the connection variables, which are far more difficult to constrain within the spin foam path integral than the bi-vectors. 

In this work, we shall discuss a phase space corresponding to a discretised Plebanski theory, in particular we shall develop discrete versions of the primary and secondary simplicity constraints. One motivation is that the construction of a phase space path integral for the discrete theory, as suggested in \cite{sergei}, clearly needs such an analysis. In particular, for the phase space path integral measure, one needs to find the symplectic structure on the reduced phase space. 

Another reason arises upon finding that current spin foam models encode a boundary Hilbert space \cite{sfbdy,sergei2} that is related to the Hilbert space of loop quantum gravity restricted to a particular graph (i.e. a discrete version of loop quantum gravity).  One should examine, however, the discussion in \cite{sergei} expressing an alternative viewpoint.  Clearly, this boundary Hilbert space should be related to a quantisation of a phase space reduced by the simplicity constraints. As was already pointed out in \cite{dr1}, once one imposes {\it all} the simplicity constraints, the reduced phase space is much smaller than the phase space corresponding to a discretised version of loop quantum gravity. It is important to understand this discrepancy in more detail and we shall comment on this issue in the course of this work. Also, if one were to derive a spin foam model from a canonical quantisation, following a standard implementation of the (second class) simplicity constraints, one would improve the reliability of the spin foam approach. The standard way to enforce second class constraints is to solve them classically.  Therefore, we need a detailed knowledge of the reduced phase space.  A canonical realisation of the discrete theory in the Plebanski formulation and its subsquent quantisation would also enlighten the connections between spin foam models and loop quantum gravity.

A canonical analysis of Plebanski theory in the continuum is provided by \cite{henneaux}. The analysis of the discrete case is usually much more subtle. There is no guarantee that the processes of Legendre transformation and discretisation commute. In particular, discretisation breaks diffeomorphism symmetry \cite{gambinipullin,diffeodiscrete}. The Hamiltonian and diffeomorphism constraints, which determine the canonical dynamics in the continuum, follow from the invariance of the action under diffeomorphisms.  Therefore, the breaking of diffeomorphism symmetry in the  discrete setting alters  the concepts of the constraints; for more details, see  \cite{gambinipullin,diffeodiscrete}. This is one of the reasons why so far there is not a consistent (i.e. first class) set of discretised Hamiltonian and diffeomorphism constraints for 4d gravity. Naturally, this affects the computation of the secondary simplicity constraints, which in the continuum are determined by the time evolution of the primary constraints.  More precisely, they are defined as the Poisson brackets between these primary constraints and the Hamilltonian constraints. Thus, we cannot derive the secondary constraints along conventional lines.\footnote{A way out is to consider the canonical analysis of the discrete action as is done in \cite{gambinipullin,diffeodiscrete}. The act of discretisation breaks the symmetries  of the  action and results in the replacement of constraints by pseudo-constraints. These pseudo-constraints are equations of motion, in which the canonical data of two consecutive time steps are only very weakly coupled to each other (Remember that true constraints are equations of motion, in which the canonical data of consecutive time steps are not coupled to each other at all.  True constraints concern only the data of a single time step.) For discrete gravity one can typically only obtain an implicit version of these pseudo-constraints \cite{diffeodiscrete}. Moreover, such an analysis has so far not been performed for the Plebanski formulation.}

To sidestep such difficulties, we proceed along a different route that was also followed in \cite{zapata,immirzi,dr1}. The main task is to identify a set of constraints on the phase space data (bi-vectors and holonomies of the spin connections) that spares only geometric configurations.  In other words, the bi-vectors should be constructible from tetrads, while the spatial part of the spin connection should be the Levi-Civita connection. Such constraints were proposed in \cite{zapata,immirzi,dr1}. 
However, the works \cite{zapata,immirzi} did not analyse the reduced phase space, which has now been performed in \cite{dr1}.  There, it was found that reducing the phase space does not correspond to just replacing the gauge group $\SO(4)$ with $\SO(3)$, which is what one might first expect if one considers the primary simplicity constraints only. 
Rather, if one takes all the constraints into account, one obtains a much smaller phase space.\footnote{This might be related to the difficulties encountered in \cite{immirzi} in finding a real symplectic structure starting with the complex phase space of the self-dual Ashtekar formulation \cite{ashtekar}. Indeed, we shall find a form of the simplicity constraints resembling the reality conditons for the complex-valued Ashtekar variables.} 
In this work, we shall unravel the underlying mechanism. The basic idea is the following.

\vspace{0.5cm}

The (canonical) Plebanski formulation starts from $\so(4)$-valued bi-vectors $X_{ij}{}^A$ and $\SO(4)$-valued holonomies $M_{ij}{}^{AB}$, both of which are associated to the triangles $\{ij\}$ of the 3d triangulation.  These variables are subject to the diagonal and cross simplicity constraints -- which correspond to the primary simplicity constraints --  as well as the so-called edge simplicity constraints \cite{dr1} -- which correspond to the secondary simplicity constraints. Additionally, there are the (first class) Gau\ss~constraints arising from the $\SO(4)$ gauge symmetry.

The diagonal simplicity and cross simplicity constraints are well known from the covariant discretisation of the Plebanski formulation \cite{mikedep}. The geometrical meaning of these constraints transpires from the decomposition of $\so(4)$ into self-dual and anti-self-dual sectors: $\so(4)=\su(2)_+\oplus \su(2)_-$. The diagonal and cross simplicity constraints impose that areas and 3d-dihedral angles constructed in the self-dual sector coincide with those constructed in the anti-self-dual sector. As these quantities allow one to reconstruct the 3d geometry of tetrahedra, one can summarise these constraints as requiring that \lq left-handed geometry equals right-handed geometry'  for each tetrahedron, as in the title of \cite{reisenberger}. One could therefore assume that the imposition of the full set of constraints determines the variables in one of the two chiral sectors as functions on the other sector.

Indeed the secondary (or edge) simplicity constraints may also be cast in a form which imposes the equality of left-handed and right-handed quantities.  
But to be able to incorporate all these equalities, conditions have to be satisfied involving quantities of one and the same chiral sector.
Importantly, they actually incorporate conditions among quantities within one chiral sector.  These are the gluing constraints \cite{ds,dr1} that ensure that the tetrahedra can be consistently glued along their common triangles.  The gluing constraints are implied by the secondary simplicity constraints for the following reason. The secondary simplicity constraints require that the spatial part of the $\SO(4)$ connection is of Levi-Civita type.  Hence, it is determined by the spatial geometry, i.e. the bi-vectors. (In complex Ashtekar variables, which are related to the variables occurring in the self-dual Plebanski formulation, these constraints are the reality conditions stating that the real part of the Ashtekar connection is the Levi-Civita connection.) However, the spatial Levi-Civita connection, i.e. the rotations between the triads associated to neighboring tetrahedra,  can be only determined if one is able to consistently glue tetrahedra together. Therefore, the gluing conditions must hold.  

Furthermore, the time components of the $\SO(4)$ connection encode the extrinsic curvature, which within a simplicial context is assimilated into the 4d-dihedral angles between neighboring tetrahedra. As we shall notice in the main text, the definition of the 4d-dihedral angles, using bi-vectors and holonomies, is somewhat abstruse. The gluing conditions ensure that the 4d-dihedral angles are unambiguously determined. The remaining part of the secondary simplicity constraints ensure that the 4d-dihedral angles constructed in the left-handed sector match those constructed in the right-handed sector, that is, that the self-dual and anti-self-dual geometries coincide.

\vspace{0.5cm}

In Section \ref{can}, we shall provide the essential foundations of a canonical formulation of discretised Plebanski theory.  We shall proceed, in Section \ref{gvar}, by invoking time-gauge and deriving a set of constraints that ensure geometricity. In that setting, there is a nice separation of duties.  One can clearly see that the primary simplicity constraints restrict the bi-vectors while the secondary constraints give conditions on the connection variables. To be able to satisfy all these secondary conditions further relations involving the bi-vectors must hold.\footnote{In the discrete setting, the distinction between degrees of freedom encoded in the bi-vectors and degrees of freedom encoded in the connection or holonomies is not as clear-cut as in the continuum.  Firstly, the definition of the discrete bi-vectors as continuum quantities integrated over the triangles in the triangulation involves the connection, see for instance \cite{thomasqsd}. Furthermore we shall adopt here a doubling of phase space variables, that is to every triangle we associate two bi-vectors and two holonomies. Between these pairs so-called matching conditions have to hold, that also mix the holonomies and the bi-vectors. See also the discussion in \cite{twisted}.}

 Following that, in Section \ref{ginv},  we shall introduce gauge-invariant variables, namely the areas, 2d-, 3d-, and 4d-dihedral angles and analyse the implications of the constraints, which were derived in the previous section, for these gauge-invariant variables. Moreover, we shall show that the diagonal, cross and edge simplicity constraints, as formulated in \cite{dr1,zapata}, follow from the gauge-variant form of the constraints developed in Section \ref{gvar}. This proves the sufficiency of the gauge-variant primary and secondary constraints,  since \cite{dr1} contains a sufficiency proof of the diagonal, cross and edge simplicity constraints (provided the configurations are non-degenerate). 
In section \ref{deg} we shall point out the importance of non--degeneracy conditions. We shall see that without these conditions certain formulations of the simplicity constraints allow for a $\SO(3)$--$BF$ like sector of non--geometric solutions, whose dimension is larger than the one describing geometric configurations.  We end with a discussion in Section \ref{disc}.

%%%%%%%%%%%%%%
% Continuum to discrete
%%%%%%%%%%%%%%

\section{Passing from the continuum to the simplicial}
\label{can}

In this section, we shall summarise the necessary fundamental facets of the discrete canonical formulation of $4d$ $BF$ and $4d$ Plebanski theory \cite{zapata,dr1}.

The action in the continuum is:
\be\label{condis01}
\cS_{Pleb,\cM}[X, w, \phi] = \cS_{\gamma BF}[X, w] + \cS_{simp}[X, \phi] = \int_\cM \left[ \big(X^A+ \frac{1}{\gamma}\star X^A\big) \wedge F[w]^A + \frac{1}{2}\phi^{AB}\;X^A \wedge X^B\right],
\ee
where $X$ is the bi-vector, $w$ is the spin connection, $F$ is its curvature, and $\gamma$ is the Barbero-Immirzi parameter.  The index $A=a\bar{a}$ is anti-symmetric in $a$ and $\bar{a}$, with $a,\bar{a}\in\{0,\ldots,3\}$, while $\wedge$ stands for the wedge product on forms.  The first term in the action describes a topological theory known as $BF$ theory. The Lagrange multiplier $\phi^{AB}$ imposes the simplicity constraints:
\be\label{condis02}
X^A \wedge X^B = V\; \epsilon^{AB} ,
\ee
where $V = \frac{1}{4!}\epsilon^{AB}\; X^{A}\wedge X^B$.  The non-degenerate class of solutions to these constraints  falls into two sectors.  More precisely, if $\tilde{V} := \frac{1}{6}\epsilon_{\lambda\mu\nu\rho}\; \epsilon^{AB}\; X_{\lambda\mu}^A\; X_{\nu\rho}^B \neq 0$, then
\be\label{condis03}
X^A = \left\{
\ba{l}
\pm(e\wedge e)^A,\\[0.1cm]
\pm\star(e\wedge e)^A,
\ea\right.
\ee
where $e^a$ is the co-frame 1-form field.  These solution sets are referred to as topological and gravitational, respectively.  For $\gamma= \infty$, substituting the topological solutions back into the action \eqref{condis01} produces an action that may be written as a total divergence, while if one substitutes for the gravitational solution set, one arrives at the Palatini action for general relativity.  Of course for finite Immirzi parameter, such distinctions are not strictly valid as all the solutions give a mix of topological and gravitational actions, but one keeps the nomenclature in any case.

In the continuum canonical analysis, one finds that there are primary simplicity constraints which are functionals of the bi-vectors only. These are not preserved under time-evolution unless certain secondary simplicity constraints are imposed.  These secondary constraints are functions of both the bi-vectors and the connection variables.  For those interested, we have summarised some properties of this canonical analysis in Appendix \ref{app3}, in particular,  the form of the primary and secondary simplicity constraints.

To obtain a discrete canonical formalism we intend to pursue a hybrid continuum-simplicial theory here, which allows for a straightforward construction of a symplectic form on phase space \cite{zapata}.  To do so, we keep time as a smooth parameter and assume that  $\cM$ can be folitated with leaves $\Sigma_t$, that is,  $\cM =\Sigma\times \mathbb{R}$.  We triangulate this 3d-hypersurface with flat tetrahedra: $\Sigma \rightarrow \Delta$.  The continuum fields are replaced by distributional analogues and integrated over suitable sub-simplices.

 Let us consider the $BF$ theory alone for a moment.  The canonical form of the action on $\Delta\times \mathbb{R}$ is: 
\be\label{condis05}
\begin{split}
&S_{\gamma BF, \Delta\times \mathbb{R}}[\{X_{ij}\}, \{M_{ij}\}, \{\mathfrak{l}\} ] 
:= \int_{\mathbb{R}} dt \sum_{i,j \in \Delta}  \Bigg[ C^{ABC}\; (X_{ij} + \frac{1}{\gamma}\star X_{ij})^A \;(M_{ij}^{-1}\dot{M}_{ij})^{BC} \\[0.2cm]
&\phantom{xxxxxxxxxxxxxxxxxxx} +({}^{1}\mathfrak{l}_{ij}){}^{A}\; {}^{1}\mathfrak{C}_{ij}{}^{A} + ({}^{2}\mathfrak{l}_{ij}){}^{AB}\;{}^{2}\mathfrak{C}_{ij}{}^{AB}+  (\mathfrak{l}_i)^A\;  \mathfrak{C}_i{}^A +( \mathfrak{l}_{edge})^{AB}\; \mathfrak{C}_{edge}{}^{AB}\Bigg],
\end{split}
\ee
where $i,j$ label tetrahedra.  Thus, the subscript $\{ij\}$ picks out a triangle (since they are each shared by two tetrahedra).  $X_{ij}$ and $M_{ij}$ are the bi-vector field and the parallel transport matrix between tetrahedra,  respectively. The tensor $C^{ABC}$ is the structure constant of $so(4)$, while $\{\mathfrak{l}\}$ is the set of Lagrange multipliers enforcing the following constraints:
\be\label{condis06}
\ba{lcl}
{}^{1}\mathfrak{C}_{ij}{}^{A} &:=& X_{ji}{}^A + M_{ji}{}^{AB} \; X_{ij}{}^B, \\[0.2cm]
{}^{2}\mathfrak{C}_{ij}{}^{AB} &:=&  (M_{ij}\; M_{ji}){}^{AB} - \delta^{AB}, \\[0.2cm]
\mathfrak{C}_{i}{}^A &:=& \sum_{j} X_{ij}{}^A,\\[0.2cm]
\mathfrak{C}_{edge}{}^{AB} &:= & (M_{i_1i_2}M_{i_1i_2}\dots M_{i_ni_1}){}^{AB}  ,
\ea
\ee
where $\{i_ki_{k+1}\}$ are the triangles sharing the edge. To each flat tetrahedron, we associate a reference frame. Therefore, a priori, there are two bi-vectors associated to a triangle: $X_{ij}$ and $X_{ji}$.  The first constraint states that they are related by an $\SO(4)$ transformation (that which parallel transports between the reference frames of the tetrahedra $\{i\}$ and $\{j\}$).  The second constraint states that the $\SO(4)$ transformation from tetrahedron $\{i\}$ to tetrahedron $\{j\}$ is the inverse of that from $\{j\}$ to $\{i\}$.  Thus, the first two constraints arise in the discrete formulation where we assign two pairs of canonical variables per triangle $X_{ij},\, M_{ij}$ and $X_{ji},\, M_{ji}$.  They ensure that these two pairs encode the same data (by expressing $X_{ji},M_{ji}$ as functions of $X_{ij},M_{ij}$). Thus, we shall refer to them collectively as \lq matching constraints'. The final two constraints are the closure (or Gau\ss) and flatness constraints of lattice $BF$ theory, respectively. 

We shall dispense with the flatness constraint since we shall not deal with the dynamical properties of gravity, let alone $BF$ theory.  We shall keep the closure constraint, since it is the lattice analogue of the Gau\ss~constraint that occurs in the canonical analysis of the Plebanski action \eqref{condis04a}.  It generates $\SO(4)$ gauge transformations.

It is convenient to separate the degrees of freedom into self-dual and anti-self-dual components using the Hodge-star operator on $\so(4)$. More details can be found in Appendix \ref{conv}.  The first term in the action tells us about the canonical commutation relations in the initial phase space, $\cP_{\Delta}$, upon which the constraints must act.  The canonically conjugate  variables are: $(M_{ij\pm}, X_{ij\pm})$, with commutations relations:
\be\label{condis07}
\ba{rcl}
\{X_{ij\pm}{}^A, M_{ij\pm}{}^{BD}\} & = & \dfrac{\gamma}{\gamma \pm 1} C^{ABC}\; M_{ij\pm}{}^{CD},\\[0.3cm]
\{X_{ij\pm}{}^A, M_{ji\pm}{}^{DB}\} & = & \dfrac{\gamma}{\gamma \pm 1} C^{ABC}\; M_{ji\pm}{}^{DC},\\[0.3cm]
\{X_{ij\pm}{}^A, X_{ij\pm}{}^{B}\} & = & \dfrac{\gamma}{\gamma \pm 1} C^{ABC}\; X_{ij\pm}{}^{C},\\[0.4cm]
\{M_{ij\pm}{}^{AB}, M_{ij\pm}{}^{CD}\} & = & 0.
\ea
\ee
The third relation is somewhat surprising, since it does not follow directly from the action, but it is necessary so that the commutators satisfy the Jacobi identity \cite{ashzap, thomasqsd}.  The splitting into self-dual and anti-self-dual is respected by the symplectic structure, i.e. for any two combinations $Y_+$ and $Z_-$ of variables from the self-dual and anti-self-dual sectors respectively, we have $\{Y_+,Z_-\}=0$.

\vspace{0.5cm}

\noindent Our aim is then to formulate a minimal set of constraints on the phase space $\cP_\Delta$, which remove non-geometrical canonical configurations.  There are several ways to test whether such constraints do the job.  But we shall call the constraints satisfactory if they allow us to reconstruct uniquely a vector for each edge of the 3d-hypersurface $\Delta$.  In other words, one can construct a co-frame for $\Delta$ from the $X_{ij}$ and $M_{ij}$. This is exactly what the simplicity constraints accomplish in the continuum.   Moreover, we might take some inspiration from the continuum theory and formulate the constraints as functions with a quadratic $X$ dependence.  Indeed, we adopted this form of the constraints in \cite{dr1}:
\be\label{condis08}
\ba{lclcl}
{}^{}\mathfrak{S}_{ij} & := &  X_{ij+}\cdot X_{ij+} - X_{ij-}\cdot X_{ij-},\\[0.2cm]
{}^{}\mathfrak{S}_{ijk} & := &  X_{ij+}\cdot X_{ik+} - X_{ij-}\cdot X_{ik-},\\[0.2cm]
{}^{}\mathfrak{S}_{ijkl} &:=&  X_{ik+}\cdot (M_{ij+}X_{jl+}) - X_{ik-}\cdot (M_{ij-}X_{jl-}),
 \ea
\ee
For the last constraints the triangles $\{ij\},\{ik\},\{jl\}$ must share an edge. These constraints are known as the diagonal, cross, and edge simplicity constraints.  Although, we do not have a lattice Hamiltonian, we refer to the first two as primary, since they are independent of $M_{ij}$, while we refer to the final one as a secondary constraint (just as in the continuum setting, see \eqref{condis04a} and \eqref{condis04b}).  All three relate self-dual and anti-self-dual phase space parameters.  

 In \cite{dr1}, we showed that provided we insist on a non-degeneracy condition,\footnote{It will turn out that these non-degeneracy conditions are crucial. In fact,  the conditions do not only require 3d non-degeneracy but also that the normals of neighboring tetrahedra are non-parallel. In the particular case, where the 3d-hypersurface $\Delta$ is the boundary triangulation of a 4-simplex, these conditions ensure 4d non-degeneracy.} the constraint set consisting of the matching constraints, the Gau\ss~constraints as well as the diagonal, cross and edge simplicity constraints    $\{ {}^{1}\mathfrak{C}_{ij}, {}^{2}\mathfrak{C}_{ij},\mathfrak{C}_{i},  \mathfrak{S}_{ij}, \mathfrak{S}_{ijk}, \mathfrak{S}_{ijkl} \}$ allows one to construct an assignment of vectors $y_i(e)$ to the edges $e$ of the triangulation. Here, the sub-index $i$ indicates that $y_i(e)$ is expressed in the frame of the tetrahedron $\{i\}$. The quadratic form of the constraints \eqref{condis08} allows for two possible sets.   One assignment is such that the bi-vectors $X_{ij}$ can be written (modulo a sign) as the Hodge-duals of wedge products of vectors associated to two of the edges $e,e'$ of the triangle $\{ij\}$:
 \be\label{condis09b}
 X_{ij}{}^A=\pm\star(y_i(e) \wedge y_i(e'))^A .
 \ee
 Moreover, these edge-vectors are consistently transported between the frames of adjacent tetrahedra, that is $y_j{}^a(e)= M_{ji}{}^{ab}\; y_i{}^b(e)$, where $M^{ab}$ is the rotation matrix in the vector representation corresponding to the matrix $M^{AB}$ in bi-vector representation, see Appendix \ref{reps}. This latter property is ensured by the secondary or edge simplicity constraints and corresponds to the continuum condition that the spatial part of the spin connection is the  Levi-Civita connection, i.e. it is determined by the spatial metric.
 
In this so-called gravitational sector,  the set of canonical variables $\{X_{ij},M_{ij}\}$  determines uniquely the assignment (\ref{condis09b}) of vectors to edges  modulo an orientation reversal for all vectors $y_i(e) \mapsto -y_i(e)$.

As we said, the constraints \eqref{condis08} allow for solutions in the so-called topological sector, which are of the form:
\be\label{topi}
X_{ij}{}^A=\pm(y_i(e) \wedge y_i(e'))^A .
\ee
These can be excluded using any one of two methods.  On might replace some of the quadratic constraints with constraints cubic in the bi-vectors or one may add discrete conditions which fix the sign of certain quantities.  This will be explained in more detail in Section \ref{relcon}.

%%%%%%%%%%%%%%
% Gauge variant
%%%%%%%%%%%%%%

\section{Gauge-variant analysis}
\label{gvar}

The constraints  \eqref{condis08} are scalar quantities and moreover, they are quadratic in the bi-vectors. Alternatively, the primary simplicity constraints can be recast in a format, which is linear in the bi-vectors. These conditions ensure that given the four bi-vectors $X_{ij}$ (varying $\{j\}$), one can reconstruct edge-vectors for the tetrahedron $\{i\}$. Thus, they have the advantage over the quadratic constraints \eqref{condis08} that they kill the topological sector. For this reason, these constraints have been applied in the more recent construction of spin foam models \cite{epr,fk, livs}. This reformulation of the constraints requires an auxiliary vector, which if fixed to $(1,0,0,0)$, leads to what is known as the time-gauge. We shall adopt the time-gauge in our analysis and later detail how to move out of this gauge.

If several tetrahedra share a given edge in the $3d$-hypersurface, then we have also to guarantee that the vectors reconstructed in the different adjacent tetrahedra for that edge are identical, that is they are consistently parallel transported into each other.  The secondary constraints, constructed below, see to this.  They are the gauge-variant equivalent of the edge simplicity constraints and thus restrict the parallel transport matrices. As we shall derive later, they lead to constraints involving the bi-vectors of neighboring tetrahedra.

\subsection{Formulation of primary constraints}
\label{spri}

Consider a tetrahedron $\{i\}$.  Should one construct the bi-vectors as functions of the edge-vectors according to \eqref{condis09b}, then together they span only a 3d space, as a tetrahedron is a 3d object. Hence, for a geometrical configuration, there exists a four-vector $n_i$ for every tetrahedron $\{i\}$  such that:
\be\label{pri00}
n_i{}^a\; (\star X_{ij})^{a\bar{a}} \approx 0,
\ee
for all four bi-vectors $X_{ij}$ in this tetrahedron. From now on, we shall refer to these gauge-variant conditions (especially in time-gauge, see below) as primary simplicity constraints. Note that this condition excludes the topological sector of solutions \eqref{topi} (for non-degenerate tetrahedra). To see this, one should remember that the Hodge-dual $\star X$ of a simple bi-vector $X$ spans a 2d plane orthogonal to the plane spanned by $X$. Therefore, for a topological configuration,  planes defined by the Hodge-duals of bi-vectors of the form (\ref{topi}) would share one direction, but three of these planes would span 4d space.

On the other hand one can show \cite{fk} that bi-vectors satisfying the condition (\ref{pri00}) also satisfy the diagonal and cross simplicity constraints.

Were one to perform a gauge transformation on the reference frame of the tetrahedron $i$, the vector $n_i$ would be rotated.  We shall fix some of this freedom by choosing $n_i =e_t := (1,0,0,0)$ for every tetrahedron.  This is known as time-gauge, and it yields a particularly tidy form of the simplicity constraints. We can formulate the primary constraints in the following manner: 
\be\label{pri01}
\mathfrak{s}_{ij}{}^r   :=    (e_t)^a\; \epsilon^{ar}{}^{B}\;X_{ij}{}^B   =  X_{ij+}{}^{0r} - X_{ij-}{}^{0r},\\
\ee
where in this and subsequent formulae we shall reserve $r,s,t$ to denote spatial indices taking values in $\{1,2,3\}$.

 Notably, the constraints become linear in the $X$ variables but at the expense of introducing an explicit normal vector to each tetrahedron in the spatial hypersurface.   A continuum version of this formalism has been developed in \cite{steffen}.

\subsection{Formulation of secondary constraints}
\label{ssec}

As stated earlier, now that one has reconstructed edge-vectors for each tetrahedron independently, one must ensure that these vectors agree for coincident edges.  Thus, one must conceive of constraints on the connection degrees of freedom that map between the vector spaces associated to different tetrahedra. Let us consider two tetrahedra $\{i\}$, $\{j\}$ with a shared face $\{ij\}$.  Their reference frames have bases $\di e^a$ and  $\dj e^a$, respectively ($a = 0,1,2,3$).  Moreover, we shall denote vectors by $ x =   x_i{}^a \di e^a$.  There exists an othogonal transformation between the two vector spaces:  $\dj e^a = \di e^b \;  M_{ji}{}^{ba}$.   Therefore, the components change like $ x_j{}^a = M_{ji}{}^{ab}\;  x_i{}^b$.

We are interested in the space of bi-vectors.  We denote a generic bi-vector by $ X_i =  X_i{}^A\;(\di e\wedge \di e)^A$,  and $(\di e\wedge \di e)^A = \di e^{a}\wedge \di e^{\bar{a}}$.  The transformation between vector spaces induces a transformation between bi-vector spaces:  $ X_j{}^A = M_{ji}{}^{AB} \;  X_i{}^B$.  

A priori, we are not in a position to describe this rotation in the vector representation, since we do not start with the vector degrees of freedom.  Rather, we start with the bi-vectors.   The bi-vectors associated to the triangle shared by the tetrahedra $\{i\}$ and $\{j\}$ are denoted $X_{ij}$ and $X_{ji}$.  After the primary constraints are imposed, one is able to reconstruct the edge-vectors for each tetrahedron.  We shall denote the edge-vectors associated to a particular edge in the triangle $\{ij\}$ as $y_{ijk}$ and $y_{jil}$, where the first index shall prescribe within which tetrahedron it is constructed. Put differently, we index an edge by $\{ijk\}$ by a triple of tetrahedra sharing this edge. Moreover, by $\{jil\}$ we usually refer to the same edge but expressed in the reference frame of the tetrahedron $\{j\}$. If an edge is shared by only three tetrahedra (as is the case for the boundary triangulation of a 4-simplex), one has $k=l$.  

To describe a geometric configuration, we must ensure that the three $y_{ijk}$ per triangle are rotated into the three $y_{jil}$.   This will give us three constraints on the possible form of the matrix $M_{ij}$.  

To be clear, a matrix in the vector representation is denoted by $M^{ab}$, a matrix in the bi-vector representation is denoted by $M^{AB}$, while once one moves to time-gauge, a matrix in the bi-vector representation is denoted $m^{rs}$.

Our transformation may be separated nicely into three parts according to a number of criteria: 
\be\label{sec06}
X_{ji}{}^A =  -M_{ji}{}^{AB}\; X_{ij}{}^B = - \big(\uoe M_{ji}\; \uo M_{ji}\; \us M_{ji}\big){}^{AB} \; X_{ij}{}^B
\ee
Note that $\star X_{ji} =  -M_{ji}\; \star X_{ij}$ if and only if $X_{ji} =  -M_{ji}\; X_{ij}$.  Speaking in rough terms, any rotation can be decomposed into a part which preserves a plane pointwise, a part which rotates within that plane and a part which actively rotates that plane into another.
\begin{description}
\item[Part 1 - $\us M_{ji}$:]  The first part of the $\SO(4)$ rotation preserves pointwise $\star X_{ij}$, that is:
\be\label{sec07}
 y_i{}^a \rightarrow (\us M_{ji}\;  y_i){}^a =  y_i{}^a \quad\quad \textrm{if} \quad  y_i\subset \star X_{ij}.
\ee
Of the three parts, this is the one that is not completely determined by the bi-vector data and it is related to extrinsic curvature of the 3d-hypersurface.    

\item[Part 2 - $\uo M_{ji}$:] The next part rotates the plane of the triangle $\{ij\}$ as seen in $\{i\}$ into the plane as seen in $\{j\}$, that is, it rotates $\star X_{ij}$ into $\star X_{ji}$:
\be\label{sec08}
\uo M_{ji}: X_{ij}{}^A\rightarrow (\uo M_{ji}\;  X_{ij}){}^B = -X_{ji}{}^A.
\ee
Such a rotation $M_{ji}$ can be found as function of the bi-vectors $X_{ij}$ and $X_{ji}$.

\item[Part 3 - $\uoe M_{ji}$:] Finally, we need the part which rotates the edge-vectors within the plane $\star(\uo M_{ji}X_{ij})$:
\be\label{sec09}
\uoe M_{ji} :  (\uo M_{ji}\; y_{ijk}){}^a \rightarrow (\uoe M_{ji}\;  \uo M_{ji}\; y_{ijk}){}^a = - y_{jil}{}^a,
\ee
where $\{ijk\}$ refers the edge shared by the triangles $\{ij\}$ and $\{ik\}$ and so on.  The reason for the minus sign in the final equality will become clear when we define the edge-vectors. Once again, this rotation is can be determined as a function of the bi-vectors.  
\end{description}
All together, these three parts describe the most general transformation which ensures the geometricity of the 3d-hypersurface triangulation.

\vspace{0.5cm}

\noindent  We shall now embark on making the above statements more explicit, beginning with:
\begin{description}
\item[Part 1:] We are interested in transformations which stabilise a particular bi-vector:
\be\label{sec10}
X^A \rightarrow (\us M\; X)^A = X^A
\ee
There are two such transformations which satisfy this criterion:
\be\label{sec11}
\us M^{AB} = \exp\big(\eta_1\; \hat{X}\cdot J\big)^{AB}, \quad\quad \textrm{and}\quad\quad \us N^{AB} = \exp\big(\eta_2\;  \star\!\hat{X}\cdot J\big)^{AB},
\ee
where $\hat{X}:= \frac{X}{||X||}$. The $J$ are the generators for $\SO(4)$, see Appendix \ref{reps} for precise definitions. Moreover, we are interested in the specific case that  X is simple and that the transformations preserve, in a pointwise fashion, the plane defined by $\star X$. (This means that the vectors associated to the edges of the triangle are stabilised also.)  The primary simplicity condition (in time-gauge) is given by \eqref{pri01}:
\be\label{sec12}
( e_t)^a \;\epsilon^{arB} \; X^{B} = 0.\nonumber
\ee
Equivalently, the condition states:  $X^{rs} = 0$, where $r,s  \in \{1,2,3\}$.  Thus, the transformations simplify to:
\be\label{sec13}
\us M^{AB} = \exp\big(\eta_1\; \hat{X}^{0r} J_{0r}\big)^{AB}, \quad\quad \textrm{and}\quad\quad \us N^{AB} = \exp\big(\eta_2 \; \hat{X}^{0r}\epsilon_{0r}{}^{st} J_{st}\big)^{AB}.
\ee
We can easily see that $\us M^{AB}$ preserves pointwise the plane defined by $\star X$, while it rotates the half-lines in the plane defined by $X$ through an angle $\eta_1$.  On the other hand, $\us N^{AB}$ preserves pointwise the plane defined by $X$, while it rotates the half-lines in the plane defined by $X$ through an angle $\eta_2$.  Thus, the case we are interested in is that in which $\eta_2 = 0$. 

At many points in our analysis, we split our degrees of freedom into self-dual and anti-self-dual variables:
\be\label{sec14}
\us M_{\pm}{}^{AB} = \exp\big(\eta_\pm\; \hat{X}_\pm\cdot J_\pm\big)^{AB}, \quad\quad \textrm{where}\quad\quad \eta_\pm = (\eta_1 \pm \eta_2)\frac{||X_\pm||}{||X||}.
\ee
Moving to time-gauge this yields:
\be\label{sec15}
\us m_{\pm}{}^{rs} := 2\; \us M_\pm{}^{0r0s} =  \exp\big(\pm\sqrt{2}\;\eta_\pm\; \hat {x}_\pm\cdot j_\pm\big)^{rs}, \quad\quad \textrm{where} \quad\quad 
\ba{rcl}
x_\pm{}^r &:= & 2\; X_\pm{}^{0r},\\[0.2cm]
(j_\pm{}^r)^{st} &:=& \epsilon^{rst}   := \epsilon^{0rst}.
\ea
\ee
The $\pm$ sign arises since $C_{\pm}{}^{0r0sC}\epsilon^{C0t} = \epsilon^{0rst}$.\footnote{
We should note that $\us M_\pm{}^{0r}{}_{0s}$ are not the only non-zero components of the matrix, but they are the only ones we need since we shall restrict to planes which lie in the 3d-hypersuface perpendicular to $e_t$.
}
 Imposing simplicity and the preservation condition ($\eta_2 = 0$), we arrive at the conclusion:
\be\label{sec16}
\eta_+ = \eta_- = \frac{\eta_1}{2}, \quad\quad x_+{}^r = x_-{}^{r} = X^{0r}
\ee
Thus, ultimately we have the condition that:  
\be\label{sec17}
\us\, m_+ = (\us\, m_-){}^{-1}.
\ee

To summarise, $\us M_{ji}{}^{AB}$ is a boost (in time-gauge). In other words, it is an $\SO(4)$ transformation effecting a rotation of the vector $e_t$, which is normal to the plane of the geometric triangle defined by $\star X_{ji}$. In time-gauge, this boost is mapped to two rotations $m_{ji\pm}$ around the 3-vector $x^r_+=x^r_-$.   The two rotation angles have the same absolute value but opposite signs.

\item[Part 2:] Now, we are in the position to examine the part of the transformation which rotates the plane $\star X_{ij}$ into $\star X_{ji}$.   
The transformation takes the form:\footnote{
It is perhaps instructive to remember the transformation in 3d which rotates the vector $a$ into the vector $b$:
\be\label{sec20a}
b^r = m^{rs}\; a^s, \quad\quad \textrm{means that} \quad\quad m = \exp\left((\sin^{-1}|\hat{b}\times \hat{a}|) \frac{\hat{b}\times\hat{a}}{|\hat{b}\times\hat{a}|}\cdot \epsilon\right),
\ee
where $(\epsilon_r)_{st} := \epsilon_{rst}$ and $\hat{a} := \dfrac{a}{|a|}$.
}
\be\label{sec20}
X_{ji\pm} = - \uo M_{ji\pm}\; X_{ij\pm},
\ee
where:\footnote{
The expansion of an exponential in the adjoint representation is:
\be\label{sec19}
\exp(\mu_\pm \hat{E}_{\pm} \cdot J_\pm)^{AB} = \cos (\sqrt{2}\mu_\pm) P_{\pm}{}^{AB} + (1 - \cos (\sqrt{2}\mu_\pm)) \hat{E}_\pm{}^A\hat{E}_\pm{}^B \pm \frac{1}{\sqrt{2}}\sin (\sqrt{2}\mu_\pm) E_\pm{}^P C_\pm{}^{PAB}
\ee
}
\be\label{sec21}
 \uo M_{ji\pm}{}^{AB} := \exp \left[-\frac{1}{\sqrt{2}}\left(\sin^{-1}\frac{||\hat{X}_{ji\pm} \times \hat{X}_{ij\pm}||}{\sqrt{2}}\right) \frac{\star(\hat{X}_{ji\pm}\times \hat{X}_{ij\pm})}{||\hat{X}_{ji\pm} \times \hat{X}_{ij\pm}||}\cdot J_{\pm}\right]^{AB}.
\ee
The transformation satisfying (\ref{sec20}) is not unique, as we can for instance multiply from the right any rotation leaving $X_{ij}$ invariant. However, this will not affect the arguments for the derivation of the secondary simplicity constraints we are going to make.

In time-gauge (\ref{sec21}) takes the form:
\be\label{sec22}
\uo\, m_{ji\pm}{}^{rs} := 2\; (\uo M_{ji\pm}){}^{0r0s} = \exp \left[ -\left(\sin^{-1}|\hat{x}_{ji\pm} \times \hat{x}_{ij\pm}|\right) \frac{(\hat{x}_{ji\pm}\times \hat{x}_{ij\pm})}{|\hat{x}_{ji\pm} \times \hat{x}_{ij\pm}|}\cdot j_{\pm}\right]^{rs}
\ee
Thus, here we can see that upon imposition of the simplicity constraint:
\be\label{sec23}
\uo\, m_{ji+} = \uo\, m_{ji-}  \quad .
\ee
In general proper rotations $M$, i.e. transformations leaving $(1,0,0,0)$ invariant, are mapped in time-gauge to two rotations $m_{\pm}$ with $m_+=m_-$.

\item[Part 3:] Ultimately, while this is enough to ensure that the triangles lie in the same plane, it is not enough to ensure that their edge-vectors are identified.  For this, we need a rotation in the plane $(\uo M_{ij}\; {\star X_{ij}})$.  To construct it explicitly, one needs to know the normalised edge-vectors.  In this gauge-fixed context, the $\hat{x}$ can be viewed as the inward-pointing 3d-normals to the triangles within the hypersurface. Therefore, if the triangles $\{ij\}$ and $\{ik\}$ share an edge $\{ijk\}$, one can describe the edge-vector using the cross product between the two normals:
\be\label{sec24}
\hat{y}_{ijk\pm}{}^r  := \frac{(\hat{x}_{ij\pm}\times \hat{x}_{ik\pm})^r}{|\hat{x}_{ij\pm}\times \hat{x}_{ik\pm}|} 
\quad\quad\textrm{and}\quad\quad 
\hat{y}_{jil\pm}{}^r  := \frac{(\hat{x}_{ji\pm}\times \hat{x}_{jl\pm})^r}{|\hat{x}_{ji\pm}\times \hat{x}_{jl\pm}|},
\ee
In addition, we have already rotated the plane in which $y_{i,jk}$ is contained:
\be\label{sec25}
\hat{y}_{ijk}{}^r \rightarrow  (\uo\, m_{ji\pm} \;\hat{y}_{ijk\pm}){}^{r}  
= \frac{ ( \uo\, m_{ji\pm}\; \hat{x}_{ij\pm} \times \uo\, m_{ji\pm}\; \hat{x}_{ik\pm})^r} {|\hat{x}_{ij\pm} \times \hat{x}_{ik\pm}|} = -\frac{(\hat{x}_{ji\pm} \times \uo\, m_{ji\pm}\; \hat{x}_{ik\pm})^r}{|\hat{x}_{ji\pm}\times \hat{x}_{ik\pm}|}.
\ee
Thus, the rotation in the plane which maps $(\uo\, m_{ji}\;y_{ijk})$ into $(-y_{jil})$ is:
\be\label{sec26}
\uoe\, m_{ji, lk\pm}{}^{rs} :=   \exp \left[ -\left(\sin^{-1}|\hat{y}_{jil\pm} \times \uo\, m_{ji\pm}\;\hat{y}_{ijk\pm}|\right) \frac{(\hat{y}_{jil\pm} \times \uo\, m_{ji\pm}\;\hat{y}_{ijk\pm})}{|\hat{y}_{jil\pm} \times \uo\, m_{ji\pm}\;\hat{y}_{ijk\pm}|}\cdot j_{\pm}\right]^{rs}.
\ee
The reason for the minus sign is due to the fact that edge-vectors are defined as the cross-product of the two normals in an ordering which necessitates its presence.\footnote{In addition, it is rather easy to see what the gauge-non-fixed expression is:
\be\label{sec27}
\uoe\, M_{ji, lk\pm}{}^{AB} :=   \exp \left[ -\frac{1}{\sqrt{2}}\left(\sin^{-1}\frac{||\hat{Y}_{jil\pm} \times \uo\, M_{ji\pm}\hat{Y}_{ijk\pm}||}{\sqrt{2}}\right) \frac{\star(\hat{Y}_{jil\pm} \times \uo\, M_{ji\pm}\hat{Y}_{ijk\pm})}{||\hat{Y}_{jil\pm} \times \uo\, M_{ji\pm}\hat{Y}_{ijk\pm}||}\cdot J_{\pm}\right]^{AB},
\ee
where $\hat{Y}_{ijk\pm}{}^A := \dfrac{(\hat{X}_{ij\pm}\times \hat{X}_{ik\pm})^A}{||\hat{X}_{ij\pm}\times \hat{X}_{ik\pm}||}$ and so forth.} Once again, we understand that once simplicity is imposed:
\be\label{sec28}
\uoe\, m_{ji,lk+} = \uoe\, m_{ji,lk-}.
\ee
The important point to notice about this part is that for a given triangle $\{ij\}$ there are three ways to construct this rotation, which correspond to which pair of edge-vectors  $y_{ijk},\, y_{jil}$ we choose.  Obviously, these should all be equal since there is only one rotation which rotates the triangles into one another. We shall comment on this point shortly.\end{description}

\vspace{0.5cm}

With the explicit form of the three rotations at hand, we can now determine the form of the secondary simplicity constraints. For every triangle $\{ij\}$, these must ensure that the rotation $M_{ji}$ consistently maps the three edge-vectors of the triangle $\{ij\}$ from the reference frame $\{i\}$ to the reference frame $\{j\}$. We constructed $\uoe\, M_{ji} \uo \, M_{ji}$ exactly such that is satisfies this requirement. Hence:
\be
( \uoe\, M_{ji} \uo \, M_{ji} )^{-1}  M_{ji}
\ee
has to be a rotation that leaves the plane $\star X_{ij}$ pointwise invariant, i.e. it  must be of the form $\us M_{ji}$.  Therefore, it satisfies (\ref{sec17}) in time-gauge.
This leads to the secondary simplicity constraints:
\be\label{sec29a}
\left[(\uoe\, m_{ji,lk+}\; \uo\, m_{ji+})^{-1}\; m_{ji+}\; (\uoe\, m_{ji,lk+}\;\uo\, m_{ji+})^{-1}\;  m_{ji-}\right]^{st} =\delta^{st},
\ee
which, since we are dealing with  $\SO(3)$ matrices, can be encoded into the form:\footnote{This form of the constraints makes it explicit that there are at most three indpendent components per triangle and edge. But it allows for an ambiguity, namely the combination of rotation matrices appearing in the square bracket could also be a rotation with an angle $\pi$. This has then to be excluded by hand and we shall implicitly understand this to be the case if one uses (\ref{sec29}).  (This problem does not appear if one uses the spin-$1/2$ representation). }
\be\label{sec29}
\mathfrak{s}_{ji,lk}{}^{r}  =  \epsilon^{rst}\left[(\uoe\, m_{ji,lk+}\; \uo\, m_{ji+})^{-1}\; m_{ji+}\; (\uoe\, m_{ji,lk+}\;\uo\, m_{ji+})^{-1}\;  m_{ji-}\right]^{st} ,
\ee
where $\uoe\, m \; \uo\, m = \textrm{function}(\{X\})$. 

The  constraints (\ref{sec29}) involve quite complicated functions of the bi-vectors compared to the primary
 simplicity constraints (\ref{pri01})). There are three constraints (with three components) per triangle $(ij)$. These fix the
  rotation $m_{ji-}$ in terms of the rotation $m_{ji+}$, the rotation built out of a vectors $x_{ij},x_{ji}$, and the rotation constructed from a pair of edge-vectors $y_{ijk},\,y_{jil}$. For the three possible pairs of edge-vectors, we obtain three conditions on the relation between $m_{ji+}$ and $m_{ji-}$.

Note that the secondary simplicity constraints resemble the reality condition for the complex Ashtekar variables \cite{ashtekar, sergeireal}, (see also \cite{immirzi} for an alternative discretisation of these conditions). These reality conditions require that the real part of the spatial pull-back of the Ashtekar connection, i.e. the sum of the self-dual and anti-self-dual connections, is equal to (twice) the spin (Levi-Civita) connection. Here, $(\uoe\, m_{ji,lk+} \, \uo\, m_{ji+})$ is a discretization of the spatial spin (Levi-Civita) connection and for every triangle may be constructed out of the bi-vectors in three different ways -- depending on the choice of edge in the triangle under consideration. To obtain an unambiguous discretisation of the spin connection, these three ways must coincide and this eventually leads to the so-called gluing conditions.

\vspace{0.5cm}
Let us summarise the phase space variables and constraints in time-gauge.
Our initial phase space parametrisation in this gauge-fixed setting is given by the quadruple $(x_{ij\pm}{}, x_{ji\pm}{},  m_{ij\pm}, m_{ji\pm})$ of data per triangle $(ij)$. These variables feature the following Poisson algebra:
\be\label{sec31}
\ba{rcl}
\{x_{ij\pm}{}^{a}, m_{ij\pm}{}^{bd}\}  & = & \pm 2 \dfrac{\gamma}{\gamma \pm 1} \epsilon^{abc} m_{ij\pm}^{cd},\\[0.3cm]
\{x_{ij\pm}{}^{a}, m_{ji\pm}{}^{db}\}  & = & \pm 2 \dfrac{\gamma}{\gamma \pm 1} \epsilon^{abc} m_{ji\pm}^{dc},\\[0.3cm]
\{x_{ij\pm}{}^{a}, x_{ij\pm}{}^{b}\}  & = & \pm 2\dfrac{\gamma}{\gamma \pm 1} \epsilon^{abc} x_{ij\pm}^{c},\\[0.3cm]
\{x_{ij\pm}{}^{a}, x_{ji\pm}{}^{b}\}  & = & 0,\\[0.3cm]
\{m_{ij\pm}{}^{ab}, m_{ij\pm}{}^{cd}\}  & = & 0.
\ea
\ee
Then, one imposes the constraint set:
\begin{eqnarray}
\label{sec32a}		{}^1\mathfrak{c}_{ij\pm}{}^r&=& x_{ji\pm}{}^r + (m_{ji\pm}\; x_{ij\pm})^r,\\[0.2cm]
\label{sec32b}		{}^2\mathfrak{c}_{ij\pm}{}^r&=& \epsilon^{rst}\;(m_{ij\pm} \;m_{ji\pm})^{st}\\[0.2cm]
\label{sec32c}		\mathfrak{c}_{i\pm}{}^r &=& \sum_{j} x_{ij\pm}{}^r,\\[0.1cm]
\label{sec32d}		\mathfrak{s}_{ij}{}^r   &= &    x_{ij+}{}^r - x_{ij-}{}^r,\\[0.2cm]
\label{sec32e}		\mathfrak{s}_{ji,lk}{}^{r} & = &\epsilon^{rst} \left[(\uoe\, m_{ji,lk+}\; \uo\, m_{ji+})^{-1}\; m_{ji+}\; (\uoe\, m_{ji,lk+}\;\uo\, m_{ji+})^{-1}\;  m_{ji-}\right]^{st} .
\end{eqnarray}
Upon imposing the above constraints, one automatically satisfies the following condition:
\be\label{sec33}
\mathfrak{g}_{ji,lk\pm}{}^{r}  =\epsilon^{rst}\;\left[  \uoe\, m_{ji,lk\pm}\,\left( \uoe\, m_{ji,l'k'\pm} \right)^{-1} \right]^{st},
\ee
that is, the rotation in the plane is independent of the pair of edges chosen to construct the transformation $m_{ji,lk}$.  It turns out that one may replace some of the secondary simplicity constraints by the constraints \eqref{sec33}.

\subsection{Relaxing time-gauge}

Time-gauge assumes that all the 4d-normals to the tetrahedra are rotated to $e_t=(1,0,0,0)$. We can also formulate the simplicity constraints for the general case by incorporating the transformations needed to rotate the normals  $n_i$ to $e_t=(1,0,0,0)$. To start, we retain the splitting of degrees of freedom into self-dual and anti-self-dual variables $x_{\pm}{}^r,m_{\pm}{}^{rs}$ using the auxiliary vector $e_t=(1,0,0,0)$, see Appendix \ref{conv}. 

The primary simplicity constraints \eqref{pri00} require that for every tetrahedron $\{i\}$ there exists a (unit) vector $n_i$ such that the Hodge-dual of the bi-vectors from this tetrahedron $\{i\}$ are normal to this vector: $n_i{}^a\;(\star X_{ij}){}^{a\bar{a}}=0$. We now consider an $\SO(4)$ rotation $N_i{}^{ab}$ rotating this vector $n_i$ to the standard one $e_t=(1,0,0,0)$:
\be\label{rel03}
n_i{}^a \rightarrow  N_{i}{}^{ab}\; n_i{}^b = e_t{}^a.
\ee
We can split $N_i=(n_{i+},n_{i-})$ into self-dual and anti-self-dual components, so that the variables:
\begin{eqnarray}\label{fini1}
\tilde x_{ij+}{}^r =n_{i+}{}^{rs}\; x_{ij+}{}^s , \quad\quad  \tilde x_{ij-}{}^r =n_{i-}{}^{rs}\; x_{ij-}{}^s
\end{eqnarray}
and:
\begin{eqnarray}\label{fini2}
\widetilde m_{ij+}{}^{ru} =n_{i+}{}^{rs}\; m_{ij+}{}^{st}\; (n_{j+}{}^{-1})^{tu} , \quad\quad  \widetilde m_{ij-}{}^{ru} =n_{i-}{}^{rs} \;m_{ij-}{}^{st}\; (n_{j-}{}^{-1})^{tu} \end{eqnarray}
are in time-gauge. Note that the rotation $N_i$ is not uniquely defined, as one can multiply  on the left by any rotation, which leaves $e_t$ invariant. Hence, $N_i$ should be understood as an element of $\SO(4)/\SO(3)$. A rotation $R$ leaving $e_t$ invariant corresponds in the $\pm$-splitting to two rotations $r_\pm$ satisfying $r_+=r_-$.  Thus, one can always choose $N_i$ such that $n_+=\textrm{id}$, for example. The primary simplicity constraints can then be formulated in the following way (see for instance \cite{fk}): for every tetrahedron $\{i\}$, there exists an $\SO(3)$ rotation $n_{i-}{}^{rs}$ such that:
\be\label{fini11}
x_{ij+} \approx n_{i-} \; x_{ij-},
\ee
for all triangles $\{ij\}$ in the tetrahedron $\{i\}$. Then, we define the variables $\tilde x_{ij \pm}$ and $\widetilde m_{ij\pm}$ according to (\ref{fini1}, \ref{fini2}). For these variables,  the secondary simplicity constraints (\ref{sec32e}) must hold, that is, we replace all the variables $x_\pm,m_\pm$ there by their tilde counterparts $\tilde x_\pm, \widetilde m_\pm$:
\be\label{07}
\ba{rcl}
 \tilde{\mathfrak{s}}_{ji,lk}{}^{r} & = &\epsilon^{rst} \left[(\uoe\, \widetilde m_{ji,lk+}\; \uo\, \widetilde m_{ji+})^{-1}\; \widetilde m_{ji+}\; (\uoe\, \widetilde m_{ji,lk+}\;\uo\, \widetilde m_{ji+})^{-1}\;  \widetilde m_{ji-}\right]^{st} .
\ea
\ee
If we use $n_+=\text{id}$, we even have $\uoe\, \widetilde m_{ji,lk+}\; \uo\, \widetilde m_{ji+}=\uoe\, m_{ji,lk+}\; \uo\, m_{ji+}$, as here it is understood that the rotations $\uoe\, m_{ji,lk+}$ and  $\uo\, m_{ji+}$  are fixed functions (\ref{sec22}, \ref{sec26}) of the variables $ x_+=\tilde x_+$. The secondary simplicity constraints can then be expressed in a similar form to (\ref{fini11}):
\be\label{fini12}
(\uoe\,  m_{ji,lk+}\; \uo\,  m_{ji+})^{-1}\;  m_{ji+}\; (\uoe\,  m_{ji,lk+}\;\uo\,  m_{ji+})^{-1}\; \approx\; n_{i-}\, m_{ij-}\,n_{j-}^{-1} .
\ee

 The matching and closure constraints  (\ref{sec32a}, \ref{sec32b}, \ref{sec32c})  can be expressed either in the
 $x_\pm,m_\pm$ variables or in the $\tilde x_\pm, \widetilde m_\pm$ variables, since these transform covariantly under rotations:
\be\label{rel05}
\ba{lclclcl}
{}^1\mathfrak{c}_{ij\pm}^r &\rightarrow& {}^1\tilde{\mathfrak{c}}_{ij\pm}{}^r&=& \tilde x_{ji\pm}{}^r + (\widetilde m_{ji\pm}\; \tilde x_{ij\pm})^r & = & (n_{i\pm}\; {}^1\mathfrak{c}_{ij\pm})^r,\\[0.2cm]
{}^2\mathfrak{c}_{ij\pm}{}^r&\rightarrow&{}^2\tilde{\mathfrak{c}}_{ij\pm}{}^r&=& \epsilon^{rst}\;(\widetilde m_{ij\pm} \;\widetilde m_{ji\pm})^{st}  &=& (n_{i\pm}\; {}^2\mathfrak{c}_{ij\pm}){}^r\\[0.2cm]
\mathfrak{c}_{i\pm}^r &\rightarrow& \tilde{\mathfrak{c}}_{i\pm}{}^r &=& \sum_{j} \tilde x_{ij\pm}{}^r &=& (n_{i\pm}\; \mathfrak{c}_{i\pm})^r.
\ea
\ee
Therefore, $ {}^1\tilde{\mathfrak{c}}_{ij\pm} \approx 0 $ if and only if ${}^1\mathfrak{c}_{ij\pm}\approx 0$ and so on.

The conditions (\ref{fini11},\ref{fini12}) involve the rotations $n_{i-}$ as auxiliary objects. To obtain constraints in the usual form, i.e. only containing phase space variables, one can expand the phase space to include the variables $n_{i-}$ and their conjugate momenta. Then, one would have to add more constraints, which basically eliminate the $n_{i-}$ and conjugate momenta as dynamical variables, so that in the end one obtains the same reduced phase space as before. Furthermore, the Gau\ss~constraints must be altered so that they also generate rotations for the $n_{i-}$ variables.   Work along these lines has been advanced in \cite{steffen, sergejyetagain} and plays an important role in the definition of the so-called projected spin networks \cite{sergejyetagain}.  A full canonical analysis is still open, however.

\section{Gauge-invariant phase space}
\label{ginv}

\subsection{Gauge-invariant variables}
\label{gaugeinvariant}

Here we shall discuss gauge-invariant variables in more detail.  We shall be interested in investigating the relations among them arising purely from their definitions as well as the conditions imposed by the primary \eqref{pri01} and secondary \eqref{sec29} constraints.

We shall define the gauge-invariant variables and discuss the relations among them using  the time-gauge. The definitions for the gauge-invariant variables are also valid if we are not in time-gauge and just use the auxiliary vector $e_t=(1,0,0,0)$ to define the $\SO(3)$ algebra (resp. group) valued variables $x_{\pm}$ (resp. $m_{\pm}$) . The relations among the gauge-invariant variables and the restrictions imposed by the constraints are of course also valid off time-gauge.

To start with, let us consider two gauge-invariant quantities that can be constructed from the data associated to one tetrahedron $\{i\}$:  the squared areas $a_{ij\pm}$ and the (outer) 3d-dihedral angles $\phi_{ijk\pm}$ between the normals $x_{ij\pm}$ and $x_{ik\pm}$ to the two triangles $\{ij\}$ and $\{ik\}$, respectively:
\be\label{defphi}
\ba{rcl}
a_{ij\pm}  &=& x_{ij\pm} \cdot x_{ij\pm}, \\[0.2cm]
\cos \phi_{ijk\pm}  &=& \Bigg[\dfrac{x_{ij}\cdot x_{ik}}{\sqrt {x_{ij}\cdot x_{ij}} \sqrt {x_{ik}\cdot x_{ik}} }\Bigg]_\pm .
\ea
\ee
There are four of the former and six of the latter for each chirality.  In each $\pm$-sector, the Gau\ss~constraints \eqref{sec32c} give four relations among the six 3d-dihedral angles per tetrahedra:
\be\label{gaussb}
\sqrt{a_{ij\pm}}+\sum_{k\neq j}  \sqrt{a_{ik\pm}}\,\cos\phi_{ijk\pm}=0 \quad .
\ee
For each tetrahedron these Gau\ss~constraints can be solved for four of the 3d-dihedral angles, so that one is left with the four area variables and two (non-opposite) 3d-dihedral angles.  This holds separately in the $+$-sector and the $-$-sector. Hence, we obtain $2\times(4+2)=12$ independent gauge-invariant variables per tetrahedron. These are sufficient to determine the left-handed and right-handed geometry of each tetrahedron.

From the 3d-angles $\phi$, one can construct the (outer) 2d-dihedral angles $\alpha$. This can be easily seen by considering directly the definition of the 2d-angles as products between the edge-vectors, for example $y_{ijk\pm}$ and $y_{ijl\pm}$:
\be\label{defalpha}
\cos\alpha_{ijkl\pm} = \left[\frac{(x_{ij}\times x_{ik})\,\cdot\,(x_{ij}\times x_{il})}{ \sqrt{(x_{ij}\times x_{ik})\cdot (x_{ij}\times x_{ik})}\sqrt{(x_{ij}\times x_{il})\cdot(x_{ij}\times x_{il}) }   } \right]_\pm \,=\, \left[  \frac{\cos \phi_{ikl}-\cos\phi_{ijk}\cos\phi_{ijl}}{\sin\phi_{ijk}\sin\phi_{ijl}}\right]_\pm.
\ee

\vspace{0.5cm}

\noindent We now turn to gauge-invariant data that require the consideration of more than one tetrahedron. Consider an edge $\{ijk\}$ in the tetrahedron $\{i\}$ and the same edge $\{jil\}$ seen from the tetrahedron $\{j\}$.\footnote{For the 3d triangulation defined by the boundary of a  4-simplex we have that $l=k$.}  We consider the angle between these two edges after parallel transporting $\{jil\}$ to the reference system of $\{i\}$:
\be\label{deftheta}
\cos\theta_{ik,jl\pm} = \left[\frac{ (x_{ij} \times x_{ik})\; m_{ij} \;(x_{ji}\times x_{jl})}{\sqrt{(x_{ij} \times x_{ik})  \cdot (x_{ij} \times x_{ik}) }\sqrt{    (x_{ji}\times x_{jl}) \cdot  (x_{ji}\times x_{jl})  }    }\right]_\pm  .
\ee
Hence, one can construct three angles, depending on the choice of edge, for every triangle. On a geometric configuration the three angles per triangle coincide and give the value of the (inner) 4d-dihedral angle between the two tetrahedra sharing this triangle (that is, the angle between the normal and reversed normal of the two neighboring tetrahedra)  \cite{dr1}. Note that this interpretation does not necessarily hold for non-geometric configurations. We shall nevertheless refer to these angles as 4d-dihedral angles.

\vspace{0.5cm}

\noindent Let us determine the dimension of the Gau\ss-reduced phase space for a generic triangulation. We started with $2\times6$ variables per triangle, hence the dimension of the initial phase space is $12\times N_t$, where $N_t$ is the number of triangles. In a 3d triangulation without boundary, every triangle is shared by two tetrahedra and every tetrahedron has four triangles. Thus, we have twice as many triangles as tetrahedra. This means that the  number of Gau\ss~constraints totals $6\times\tfrac{1}{2}\times N_t$. As these are first class constraints, we subtract  twice their number from the dimension of the initial phase space.  Therefore,  the dimension of the reduced phase space is $6\times N_t$. Remember that there are two area variables per triangle and four independent 3d-dihedral angles per tetrahedron, which amounts to four variables per triangle.\footnote{
We implicitly imposed the \lq matching' constraints (\ref{sec32a},\ref{sec32b}).  They are second class constraints, but have trivial effect on the Poisson structure, and ultimately, all one needs to do is implicitly remember that if one writes down $m_{ij\pm}$ and $x_{ji\pm}$, that they are equal to $(m_{ji\pm}){}^{-1}$ and $-m_{ji\pm}x_{ij\pm}$, respectively.
}
This leaves two further variables per triangle, in order to fully parameterise the Gau\ss-reduced phase space.  We must choose from the 4d-angles $\theta$ defined in \eqref{deftheta}, of which there are three per triangle and chiral sector.  

\vspace{0.5cm}

\noindent This suggests that these six $\theta$-angles are not independent variables. Indeed, as we shall explain soon,  they are related by terms involving the 2d-angles $\alpha$.

\vspace{0.5cm}

\begin{figure}[H]
\centering
\includegraphics[width = 10cm]{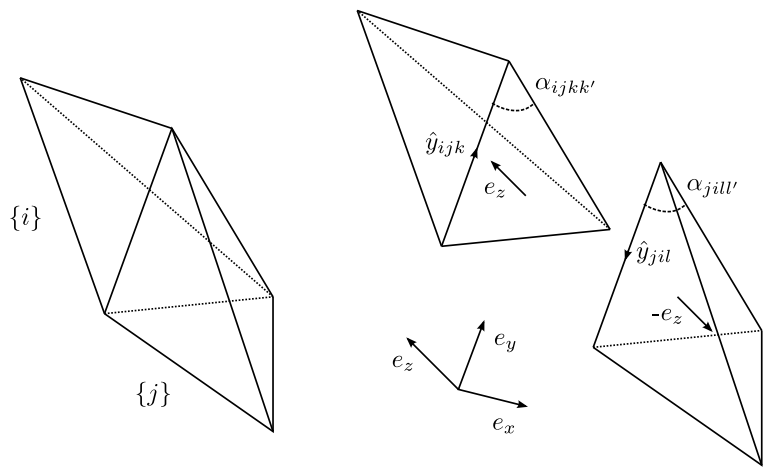}
\caption{\label{tetatet}Two tetrahedra sharing a triangle and the associated 2d dihedral angles and edge vectors.}
\end{figure}

\noindent To analyse this interdependence,  let us consider two tetrahedra $\{i\}$ and $\{j\}$, as illustrated in Fig. \ref{tetatet}.  We choose a gauge such that $x_{ij\pm}=(0,0,1)$ and $x_{ji\pm}=(0,0,-1)$. Due to the conditions $x_{ij\pm}=-m_{ij\pm}\;x_{ji\pm}$, the rotation matrices $m_{ij\pm}$ must now leave the vector $e_z=(0,0,1)$  invariant, i.e. they are $\SO(2)$ rotations. Thus, we can write  $m_{ij\pm}=R(\xi_\pm)$ where $R(\xi)$ is a (positively-oriented) rotation around $e_z$ with angle $\xi$.  The angles $\xi_\pm$ are still not gauge-invariant quantities,\footnote{These angles are also used in \cite{immirzi,valentin,twisted}.}  since there is a residual gauge freedom: one may still rotate around $e_z$ in the frames of both tetrahedra.  We can fix this residual gauge freedom by demanding, for instance, that the edge $\{ijk\}$ in the tetrahedron $\{i\}$ and the corresponding edge $\{jil\}$ in the tetrahedron $\{j\}$ are labelled by the normalised vectors $\hat y_{ijk\pm}=(0,1,0)$ and $\hat y_{jil\pm}=(0,-1,0)$, respectively. Here, we used again the definitions \eqref{sec24}:
\begin{eqnarray}
\hat y_{ijk\pm} :=\left[\frac{ x_{ij} \times x_{ik} }{  \sqrt{  (x_{ij} \times x_{ik}) \cdot (x_{ij} \times x_{ik})}} \right]_\pm \nonumber.
\end{eqnarray}
In this gauge, we obtain $m_{ij\pm}=R(\pi-\theta_{ik,jl\pm})$. Note that the definition \eqref{deftheta} for $\theta_{ik,jl\pm}$ does not fix the sign of the angle.  This sign is determined  by the equation $m_{ij\pm}=R(\pi-\theta_{ik,jl\pm})$.

All the edge-vectors associated to the triangle $\{ij\}$,  in the frames $\{i\}$ and $\{j\}$,  are in the plane spanned by $e_x=(1,0,0)$ and $e_y=(0,1,0)$. Hence, one can write:
\begin{eqnarray}
\hat y_{ijk'\pm}= R(\alpha_{ijkk'\pm})\; \hat y_{ijk\pm} , \quad\quad \textrm{and} \quad\quad \hat y_{jil'\pm}= R(\alpha_{jill'\pm})\; \hat y_{jil\pm} ,
\end{eqnarray}
where again we augment the definitions (\ref{defalpha}) of the $\alpha$-angles so as to fix their signs.

The angles $\theta_{ik,jl}$, $\theta_{ik',jl'}$ are attached to the triangle $\{ij\}$. They are computed using the pairs of edges $\{ijk\}$, $\{jil\}$ and $\{jik'\}$, $\{jil'\}$, respectively. For the relation between these angles we compute
\be
\ba{rcl}
\cos\theta_{ik',jl'\pm}&=& \hat y_{ijk'\pm}   \cdot m_{ij\pm}\; \hat y_{jil'\pm},\\[0.2cm]
&=& \left[\left( R(\alpha_{ijkk'})\;\hat  y_{ijk}\right) \cdot R(\pi- \theta_{ik,jl}) \; \left(   R(\alpha_{jill'})\; \hat y_{jil} \right) \right]_\pm,\\[0.2cm]
&=& \left[ \hat  y_{ijk}\; R( (\pi-\theta_{ik,jl}- \alpha_{ijkk'} + \alpha_{jill'})\;  \hat y_{jil} \right]_\pm .
\ea
\ee
Let us reiterate that $\alpha_{ijkk'}$ and $\alpha_{jill'}$ correspond to the angle between the same two edges of the triangle $\{ij\}$. The difference is that $\alpha_{ijkk'}$ is computed from the data of the tetrahedron $\{i\}$ whereas $\alpha_{jill'}$ is computed from the data of the tetrahedron $\{j\}$.  Let us define the sign of the angle $\theta_{ik',jl'\pm}$ in the same way as for $\theta_{ik,jl\pm}$.  We finally obtain:
\be\label{thetaalpha}
\theta_{ik',jl'\pm} = \theta_{ik,jl\pm}\,+\alpha_{ijkk'\pm} - \alpha_{jill'\pm}.
\ee
Put into words,  the $\theta$-angles for a given triangle are defined using one edge but in two reference frames.  Meanwhile, any 2d-dihedral angle is defined using two edges but in the same reference frame.  Equation \eqref{thetaalpha} above states that the difference in $\theta$-angles defined on two edges equals the difference in 2d-dihedral angles between those two edges as computed in the two reference frames.  This explains why on geometrical configurations the three $\theta$-angles per sector and triangle coincide: on these configurations the so-called gluing conditions hold \cite{ds,dr1}, requiring that the 2d-dihedral angles at a triangle $\{ij\}$ computed from the tetrahedron $\{i\}$ should coincide with those computed from $\{j\}$. For our example, the gluing conditions impose that $\alpha_{ijkk'\pm} = \alpha_{jill'\pm}$, which implies that $\theta_{ik',jl'\pm} = \theta_{ik,jl\pm}$.

\subsection{Restrictions imposed by the primary and secondary simplicity constraints}
\label{rest}

Let us examine the implications of the primary  \eqref{sec32d} and secondary  \eqref{sec32e} simplicity constraints  for the scalar variables introduced in the previous section. 

The primary simplicity constraints \eqref{sec32d} impose that $x_{ij+}=x_{ij-}$ for all triangles $\{ij\}$. We therefore conclude that any quantity, which can be computed without using the holonomies $m_{ij\pm}$, has the same value in both the self-dual and anti-self-dual sectors, that is:
\be\label{condis}
\ba{lcl}
a_{ij+}&=&a_{ij-},\\[0.2cm]
\phi_{ij+}&=&\phi_{ij-},\\[0.2cm]
\alpha_{ijk+}&=&\alpha_{ijk-}   .
\ea
\ee
The secondary simplicity constraints \eqref{sec32e} are given by: 
\be
 \left[(\uoe\, m_{ij,kl+}\; \uo\, m_{ij+})^{-1}\; m_{ij+}\; (\uoe\, m_{ij,kl+}\;\uo\, m_{ij+})^{-1}\;  m_{ij-}\right] =\text{id}.
\ee

In the gauge that we introduced in Section \ref{gaugeinvariant}, the matrices $\uoe\, m_{ij,kl+}$  and  $\uo\, m_{ij+}$ are equal to the identity matrix. Consequently, we obtain the condition:
\begin{eqnarray}
m_{ij+}=\left(m_{ij-}\right)^{-1} .
\end{eqnarray}
Since $m_{ij\pm}=R(\pi- \theta_{ik,jl \pm})$ in this gauge, we can conclude (taking into account the $2\pi$-periodicity of the angles):
\be\label{condtheta}
\theta_{ik,jl+} = -\theta_{ik,jl-}  .
\ee
Although obtained in a specific gauge, these equations hold generally as only gauge-invariant quantities are involved. In particular, we have the conditions \eqref{condtheta} not only for the pair of corresponding edges $\{ijk\}$, $\{jil\}$ but also for the other two pairs $\{ijk'\}$, $\{jil'\}$ and $\{ijk''\}$, $\{jil''\}$. Hence, we obtain three constraints per triangle:
\be\label{condtheta2}
\ba{lcl}
\theta_{ik,jl+}&=&-\theta_{ik,jl-}\\[0.2cm]
\theta_{ik',jl'+}&=&-\theta_{ik',jl'-}\\[0.2cm]
\theta_{ik'',jl''+}&=&-\theta_{ik'',jl''-}    .
\ea
\ee
Now, we are in a position to derive gluing conditions explicitly from the primary \eqref{sec32d} and secondary \eqref{sec32e} simplicity constraints. In the previous section, we derived that the $\theta$-angles for the same triangle are related by differences in the 2d-dihedral angles $\alpha$  \eqref{thetaalpha}:
\be\label{thetaalpha2}
\theta_{ik',jl'\pm} = \theta_{ik,jl\pm} + \alpha_{ijkk'\pm} - \alpha_{jill'\pm} .
\ee
If we substitute the conditions $\theta_+ = -\theta_-$ \eqref{condtheta2} and $\alpha_+ = \alpha_-$ \eqref{condis} into equation \eqref{thetaalpha2}, we obtain the conditions:
\be\label{gluingb}
\alpha_{ijkk'\pm}=\alpha_{jill'\pm}.
\ee
In other words, the 2d-dihedral angles belonging to a triangle, as computed from the two adjacent tetrahedra, have to coincide. These are the gluing conditions introduced and discussed in \cite{ds, dr1, bahrdittrichnewangle}.  Here, we derived these gluing conditions explicitly  from the primary and secondary simplicity constraints. 

 The discussion above also allows one to derive an alternative version of the gluing conditions to those given in \eqref{gluingb}, which uses the angles $\theta$ instead of the 2d-angles $\alpha$. Namely, along with the constraints \eqref{gluingb},  we also ensure the coincidence of  the 4d-dihedral angles as computed from the three different edges of the same triangle:
 \be\label{gluingc}
\theta_{ik,jl\pm} \,=\, \theta_{ik',jl'\pm}\, = \, \theta_{ik'',jl''\pm} \quad  .
\ee 
A similar relation can be derived for the 4d-dihedral angles of a (flat) 4-simplex, where these 4d-dihedral angles are computed from the 3d-dihedral angles in different ways, see \cite{ds, bahrdittrichnewangle}. Here, we generalised this relation to a phase space context, where the 4d-dihedral angles encode the extrinsic curvature. Hence, it is a free momentum variable and is therefore not determined by the 3d-dihedral angles.

Note that whereas the primary simplicity constraints solely constitute conditions between the self-dual and the anti-self-dual sectors, the secondary simplicity constraints imply more. Specifically, they not only administer conditions between the sectors,  but they also contain relations among quantities of one and the same sector.

\subsection{Relation to the diagonal, cross and edge simplicity constraints}
\label{relcon}

In this section, we shall reconsider the diagonal, cross and edge simplicity constraints \eqref{condis08}:
\be\label{osimpl}
\ba{rclcl}
0 &=&  X_{ij+}\cdot X_{ij+} - X_{ij-}\cdot X_{ij-},\\[0.2cm]
 0&=& X_{ij+}\cdot X_{ik+} - X_{ij-}\cdot X_{ik-},\\[0.2cm]
 0&=& X_{ik+}\cdot (M_{ij+}X_{jl+}) - X_{ik-}\cdot (X_{ij-}X_{jl-}), 
 \ea
\ee
and show that these constraints follow from the primary and secondary simplicity constraints (\ref{sec32d},\ref{sec32e}).

To give ourselves a solid basis, let us restate some foundational facts.  A geometrical configuration is one in which the bi-vectors are of the form $X =\star (y(e)\wedge y(e'))$, where $y(e)$ denotes an edge-vector,  and the holonomies ensure that the edge-vectors are assigned consistently across the tetrahedra.   If the matching, Gau\ss, diagonal, cross and edge simplicity constraints hold, one can prove \cite{dr1} that on non-degenerate configurations, one has two possible ways to write the bi-vectors $X^A$.   They can be consistently written either as wedge products of vectors associated to the edges $X^A=\pm (y(e)\wedge y(e'))^A$ or as the Hodge-dual of such wedge products $X^A=\pm\star (y(e)\wedge y(e'))^A$. The first case corresponds to the topological configurations, the second to the gravitational or geometrical configurations. The non-degeneracy conditions discussed in \cite{dr1} do not only mean 3d-non-degeneracy, i.e. the non-vanishing of 3d-volumes, but it also incorporate conditions pertaining to the embedding of the 3d-triangulation into the 4d-manifold. Namely, it prescibes that the normals of two neighboring tetrahedra must be non-parallel. (As an aside, remember that the normal to a tetrahedron is well-defined if the diagonal and  cross simplicity constraints, as well as the 3d-non-degeneracy condition, hold.) Indeed in section \ref{deg} we will construct non-geometric configurations satisfying the Gau\ss, diagonal, cross and edge simplicity constraints, however featuring parallel 4d normals to the tetrahedra. This shows that the condition on non-parallel normals is necessary.

Thus, if one shows that the primary and secondary simplicity constraints (\ref{sec32d},\ref{sec32e}) imply the diagonal, cross and edge simplicity constraints \eqref{osimpl}, then one has shown that these conditions are sufficient to ensure the geometricity of the triangulation for non-degenerate configurations.  Furthermore, these constraints are also necessary, since they are satisfied by all geometric configurations. In fact, a nice feature of the primary and secondary simplicity constraints is that topological configurations are automatically excluded, as we shall see below. In contrast, since the simplicity constraints \eqref{osimpl} are quadratic in the bi-vectors, they can not distinguish between the topological and gravitational sectors, which are Hodge-dual to each other. In that case, to exclude the topological sector, one must add some conditions that are cubic in the bi-vectors (or replace some quadratic constraints by cubic ones, see also the discussion in \cite{dr1}). 

Importantly, as we shall comment in Section \ref{deg}, the non-degeneracy conditions are essential if one uses the diagonal, cross and edge simplicity constraints \eqref{osimpl}.  In contradistinction, the secondary simplicity constraints \eqref{sec32e} do not to rely on the non-degeneracy conditions and therefore constitute stronger conditions on the sector of 4d-degenerate configurations.

This discussion about the equivalence of constraints (at least on the non-degenerate gravitational sector) will allow us to freely switch between the different constraint sets and to select the most convenient one in order to compute the Dirac brackets. This computation will be done in future work \cite{dr3}. To this end, we shall determine a set of independent constraints in the next section, which will also enable us to find the dimension of the reduced phase space.

\vspace{0.5cm}

In Section \ref{rest}, we derived conditions on the gauge-invariant variables from the primary and secondary simplicity constraints. We commence by showing that the diagonal, cross and edge simplicity constraints follow from these conditions.
As shown in the previous section, the primary simplicity constraints imply \eqref{condis}:
\be\label{rc01}
a_{ij+}=a_{ij-} \quad \quad \textrm{and} \quad\quad \phantom{\cos} \phi_{ijk+}= \phi_{ijk-}.\phantom{\cos}
\ee
The diagonal and cross simplicity constraints in \eqref{osimpl} are equivalent to:\footnote{We must assume non-vanishing areas. Moreover, for the rest of the discussion we shall assume 3d-non-degeneracy, i.e. the non-vanishing of areas, 3d-dihedral angles and 3d-volumes.}
\be\label{rc02}
a_{ij+}=a_{ij-} \quad \quad \textrm{and} \quad\quad \cos \phi_{ijk+}= \cos\phi_{ijk-},
\ee
which clearly follow from \eqref{rc01}. 
The equations \eqref{rc01} also imply that the orientation in the $+$-sector and $-$-sector agree. To have a similar property with the quadratic constraints, one must add an extra constraint.  For instance, one might  require:
\be
C^{ABC}\,\epsilon^{CD}\; X_{ij}{}^A\; X_{ik}{}^B\; X_{il}{}^D =0,
\ee
 which in time-gauge reduces to:
\be
 x_{ij+}\, \cdot( x_{ik+}{} \,\times\, x_{il+}{}) \,=\,   x_{ij-}\, \cdot ( x_{ik-} \,\times\, x_{il-})
\ee
This states that the oriented volume squared as computed in the two sectors coincides.

\vspace{0.5cm}

\noindent The secondary simplicity constraints imply \eqref{condtheta}: $\theta_{ij,kl+}=-\theta_{ij,kl-}$.  For a moment, consider the gauge-invariant quantity:
\be
\cos\rho_{ik,jl\pm} = \left[ \frac{ x_{ik} \cdot (m_{ij}\;  x_{jl} )}{ \sqrt{ x_{ik}\cdot x_{ik}}\;   \sqrt{ x_{jl}\cdot x_{jl}}     } \right]_\pm .
\ee
This is the 3d-dihedral angle between the triangles $\{ik\}$ and $\{jl\}$ (where the tetrahedra $\{i\}$ and $\{j\}$ share the triangle $\{ij\}$).  Furthermore, there exists a relationship among various dihedral angles:
\be\label{rho2}
\cos\theta_{ik,jl\pm} = \left[\frac{-\cos\rho_{ik,jl} - \cos \phi_{ijk} \cos  \phi_{jil}}{\sin \phi_{ijk} \sin  \phi_{jil}} \right]_\pm .
\ee
Thus, if the $\theta_{ij,kl+}=-\theta_{ij,kl-}$ and $\phi_{ijk+}=\phi_{ijk-}$ hold, then:
\be\label{rho3}
 \cos \rho_{ik,jl+} = \cos \rho_{ik,jl-}.
\ee
But the edge simplicity constraint, the last constraint in \eqref{osimpl}, appears in time-gauge as:
\be\label{edge2}
x_{ik+} \cdot (m_{ij+}  x_{jl+}) = x_{ik-}  \cdot (m_{ij-}  x_{jl}), 
\ee
which is equivalent to \eqref{rho3}.

As before, if one passes in the opposite direction, the edge simplicity constraint only implies that $\cos \theta_{ik,jl+} = \cos\theta_{ik,jl-}$.  Therefore, one must add a further condition to the edge simplicity constraints, in order to fix the (relative) sign between the 4d-dihedral angles.   Alternatively, one may consider a cubic version of the edge simplicity constraints \cite{zapata, dr1}:
\be
X_{ik}{}^A\; M_{ij}{}^{AB}\; C^{BCD}\;  X_{ji}{}^C X_{jl}{}^D= 0   ,
\ee
which in time-gauge reads:
\be
x_{ik+}\cdot \, m_{ij+} (x_{ji+}\times x_{jl+}) \,+\, x_{ik-}\cdot \, m_{ij-} (x_{ji-}\times x_{jl-}) =0  .
\ee
This constraint implies $\theta_{ij,kl+}=-\theta_{ij,kl-}$.

\subsection{Worked example: the boundary of a 4-simplex}
\label{bdysimp}

\noindent Our next task is to select an independent subset of simplicity constraints. This will facilitate a computation of the Dirac brackets in future work \cite{dr3} and determines the dimension of the phase space corresponding to geometrical configurations.  Here, we shall consider this question on the phase space reduced by the Gau\ss~constraints.  Such a reduction simplifies hugely the calculation.  Furthermore, we shall also restrict to the phase space associated to the 3d-boundary of a 4-simplex.   More complicated triangulations will be the subject of further work.

The boundary of a 4-simplex has 5 tetrahedra connected by 10 triangles.   As explained in Section \ref{gaugeinvariant}, the Gau\ss-reduced phase space is 60 dimensional and may be parametrised using gauge-invariant variables.   The 60 independent gauge-invariant variables needed to parameterise it are provided by the 20 area variables $a_{ij\pm}$, 20 3d-dihedral angles $\phi_{ijk\pm}$ (two 3d-angles, associated to non-opposite edges, per tetrahedron and chiral sector) and 20 4d-dihedral angles $\theta_{ik,jk\pm}$ (one 4d-angle per triangle and chiral sector).\footnote{Remember that $l=k$ for the boundary of the 4-simplex. Also, note that one can explicitly reconstruct the other 3d-dihedral and 4d-dihedral angles using (\ref{gaussb}, \ref{defalpha}, \ref{thetaalpha}).}

Now we move onto the constraint set.  We shall show that the maximal independent subset of diagonal, cross and edge simplicity constraints contains 40 elements.  We shall denote this set $\mathcal{C}$.  Given the parameterisation above, one can see that an independent subset of diagonal and cross simplicity constraints is given by taking all 10 diagonal simplicity constraints, and two cross simplicity constraints (acting on the angles associated to the same two non-opposite edges) per tetrahedron.  These constitute the first 20 elements of $\mathcal{C}$.

We are left only to deal with the edge simplicity constraints.  A priori, there are thirty edge simplicity constraints ($\cos\theta_{ik,jk+}=\cos\theta_{ik,jk-}$ or $\theta_{ik,jk+}=-\theta_{ik,jk-}$), three per triangle.  We shall argue that only twenty are independent.    As illustrated in Section \ref{gaugeinvariant}, we can choose one constraint per triangle in \eqref{condtheta2} to explicitly relate quantities in the $+$-sector with corresponding quantities in the $-$-sector. These make up the next 10 elements of $\mathcal{C}$.  We showed that the remaining edge simplicity constraints implicitly impose the gluing conditions \eqref{gluingb}, which in this context take the form:
\be\label{gl2}
\alpha_{ijkk'\pm}=\alpha_{jikk'\pm}.
\ee
In words,  the 2d-angles for a triangle $\{ij\}$ coincide if computed from the two neighbouring tetrahedra $\{i\}$ and $\{j\}$.  Written in this form, there are 60 such gluing constraints, of which 10 are independent.   We shall show that on the gauge-invariant phase space, with the 30 constraints already in $\mathcal{C}$ at our disposal,  one has 50 relations among the gluing constraints \eqref{gl2}.  

For a start, note that the 2d-dihedral angles may be written in terms of the 3d-dihedral angles using the identity \eqref{defalpha}.
Using the diagonal and cross simplicity constraints in $\mathcal{C}$, one has $\alpha_+ = \alpha_-$.  So if the gluing constraints hold for the $+$-sector, they will automatically hold for the $-$-sector.   We are left now with the 30 gluing constraints on the $+$-sector, three per triangle.   But remember on the gauge-invariant phase space, one has imposed the Gau\ss~constraint.  This means that the tetrahedron is closed and furthermore it implies that the three 2d-angles for one triangle some up to $\pi$. In this way,  we obtain 10 more relations among the gluing constraints. 

The final 10 relations are quite involved since they are more non-local. But they can be understood to arise in the following manner.  One may rewrite the gluing constraints as the conditions that the length of any given edge, computed in the three adjacent tetrahedra, must be unique. From these three constraints per edge only two are independent, that is we have another set of 10 relations.  These are a manifestation of the final relations among the gluing constraints. 
Furthermore, these considerations can be confirmed by computing the Jacobian of the matrix of derivatives of the constraints with respect to the gauge-invariant variables \cite{ds}.  

Thus, we have 10 independent gluing constraints, and we may pick them in a symmetric manner, say such that they simultaneously one per triangle and one per edge.  These are the last 10 constraints to go into the set $\mathcal{C}$.

There is another way to formulate these last 10 gluing constraints. The 30 2d-angles $\alpha_+$ are determined, through the relations \eqref{gaussb} and \eqref{defalpha}, by the 10 squared areas $a_+$ and the set of 10 3d-angles $\phi_+$. We have 10 independent gluing conditions between the 2d-angles. In light of the preamble, these descend to 10 independent relations between the areas and 3d-angles.  For a geometric 4-simplex, the lengths of the 10 edges are free parameters. An alternative set of free parameters is provided by the areas.\footnote{This is the case because the matrix of derivatives of the areas with respect to the lengths is invertible at generic points.} Hence, the gluing constraints fix\footnote{modulo discrete ambiguities} the values of the 10 3d-dihedral angles $\phi_+$ as functions of the 10 (squared) areas:
\be
\phi_{ijk+}=f_{ijk}(a_{lm+}) .
\ee 
Unfortunately, an explicit expression for these functions $f_{ijk}$ is not known. This problem is equivalent to expressing the ten lengths of a 4-simplex as a function of the ten areas, which however requires the solution of a polynomial of high degree with general coefficients.

In summary, we end up with the set $\mathcal{C}$ of 40 independent constraints. $\mathcal{C}$ contains the 10 diagonal simplicity constraints and 10 cross simplicity constraints involving $\pm$--pairs of two (non-opposite) 3d-angles per tetrahedron. These 20 constraints can be seen as primary constraints, ensuring the geometricity of each tetrahedron. Then, $\mathcal{C}$ contains another  20 constraints, which in a certain sense are conjugate to the first 20 constraints. We can interpret these constraints as secondary as they ensure the consistent gluing between the tetrahedra. The 20 constraints  are formed from a set of 10 edge simplicity in the form $\cos\theta_{ikjk+}=\cos\theta_{ikjk-}$ and (10 further edge simplicity in the guise of) 10 gluing constraints: $\phi_{ijk+}=f_{ijk}(a_{lm+})$.

These 40 constraints reduce the initial 60 dimensional (already Gau\ss-reduced) phase space to a 20 dimensional one. This phase space can be parametrised by the 10 areas (or lengths) of the simplex as well as 10 4d-dihedral angles associated to the triangles of this simplex. The areas and the 4d-dihedral angles will be conjugated to each other \cite{dr1,dr3}.

\section{Degenerate configurations}
\label{deg}

The diagonal, cross and edge simplicity constraints in the form (\ref{condis08}) require that certain pairs of bi-vectors (including bi-vectors transported via a holonomy) span only a three-dimensional space. In particular, for two simple bi-vectors $X_1{}^A$ and $X_2{}^A$ satisfying:
\be
\epsilon^{AB}\; X_1{}^A\; X_2{}^B=0,
\ee
one can conclude that there exist three vectors $u,v,w$ such that:
\be
X_1{}^A =(u\wedge v)^A, \quad\quad X_2{}^A=(u\wedge w)^A .
\ee
 For a proof, see for instance \cite{fk}.  
 
But the simplicity constraints in the form \eqref{condis08}  may fail to ensure geometricity on degenerate configurations. This  was also noted in \cite{dr1}, where it was pointed out that the edge simplicity constraints also allow non-geometric configurations if the 4d-normals to two adjacent tetrahedra are parallel. For instance, consider  a configuration on the boundary of a 4-simplex, where all the bi-vectors $\star X$ are orthogonal to the unit vector $(1,0,0,0)$. In other words, we are in time-gauge. Additionally, let us say that the $\SO(4)$ matrices $M^{AB}$ are, in fact, elements of its $\SO(3)$ subgroup that leaves $(1,0,0,0)$ invariant. Then, all the bi-vectors entering the simplicity conditions are orthogonal to $(1,0,0,0)$ and so any pair of these bi-vectors spans only a three-dimensional space and the diagonal, cross and edge simplicity constraints are satisfied. Importantly, however, this is not a geometric configuration.

 In fact, note that this subspace of configurations corresponds to an $\SO(3)$ $BF$ phase space.  A related space of configurations was found in a saddle point analysis for the EPR spin foam model in \cite{nottingham}.  For the example taken, namely,  the boundary of a 4-simplex,  there are 30 independent gauge-invariant quantities in this subspace of the phase space. This is the dimension of the Gau\ss-reduced $\SO(3)$ $BF$ phase space.  Moreover, it is 10 dimensions larger than the subspace containing the geometric configurations.  For the subspace containing degenerate configurations, the dihedral angles $\theta$, as defined in \eqref{deftheta} cannot be interpreted as 4d-dihedral angles. This interpretation is only valid for geometrical configurations, while the condition that these degenerate configurations satisfy is that the information they encode is confined to a 3d space.

This shows that if one uses the form \eqref{condis08} of the constraints, one must also require 4d-non-degeneracy. In particular, the normals of neighbouring tetrahedra should not be parallel. This non-degeneracy requirement appears also in every spin foam model \cite{asdf}, see in particular the discussion in \cite{cf}. Essentially, one considers there the configuration space associated to a 4-simplex: the 10 $\so(4)$-valued  bi-vectors associated to the triangles of the simplex, which satisfy the diagonal and cross simplicity constraints as well as the Gau\ss~constraints. The connection is assumed to be flat. For non-degenerate configurations, one can show that these requirements allow the reconstruction of the 4-simplex geometry.  Put differently, one can consistently assign vectors to the edges of the 4-simplex \cite{asdf}. This geometry is characterised by 10 gauge-invariant quantities. We may choose the 10 lengths or the 10 areas of the 4-simplex. For  degenerate configurations, however, one obtains a 15-dimensional space of gauge-invariant quantities, see also  \cite{cf}.

Once again, the edge simplicity constraints in the form (\ref{condis08}) fail for 4d-degenerate configurations.  We wish to point out that this is not the case for the gauge-variant form of the secondary simplicity constraints \eqref{sec29a}, which therefore constitute a stronger requirement concerning 4d-degenerate configurations.  

Also, recasting some of the edge simplicity constraints  as  gluing conditions  $\alpha_{ijkk'}=\alpha_{jikk'}$ \eqref{gluingb} restricts the 4d-degenerate configurations. The gluing conditions do not completely exclude degenerate configurations, but they do however reduce the dimensionality of the degenerate sector. In effect, one its left with degenerate geometries, while non-geometric configurations are removed.  Therefore, they could also be applicable to spin foams where they might suppress the degenerate sector by reducing its dimensionality.

Whereas for one 4-simplex, the reduction eliminates 5 dimensions and still leaves the areas (encoded in the spins associated to the triangles) as free variables, the situation is much more complicated for a larger triangulation -- both for the 3d hypersurfaces and the 4d triangulations. Then, for geometric configurations, the areas are not any more free variables. Actually, there are (non--local) constraints among the areas to ensure that they follow from a consistent assignment of edge-lengths in the triangulation (whose number is typically smaller then the number of triangles) \cite{arearegge}. The formalism of area-angle Regge calculus \cite{ds} allows one to express these constraints in a local manner by keeping the 3d-dihedral angles as free variables. Then, the gluing constraints \eqref{gluingb} together with the Gau\ss~constraints lead to restrictions on the areas of the triangulation.

To summarise, the condition of non-degeneracy plays a crucial role in the reduction from topological $BF$ theory to the gravitational sector.
For spin foam models, it would therefore be important to study whether the dynamics encoded in the amplitudes and measure passes interchangable among non-degenerate and degenerate sectors. Were the answer affirmative, then one would have to devise methods to impose non-degeneracy conditions on the measure and amplitudes.  Possibilities for this include the gluing constraints (\ref{gluingb}) and the secondary constraints in the form (\ref{sec29a}).

\section{Discussion}\label{disc}

Spin foam models attempt to provide a path integral for gravity by starting from a partition function for topological $BF$ theory and implementing simplicity constraints within the amplitudes and measure. The correct imposition of these constraints is therefore central to the entire approach. In current spin foam models, one typically deals only with the primary simplicity constraints, but it has been pointed out  \cite{sergei} that both primary {\it and} secondary simplicity constraints  are necessary in order to produce the correct expectation values. Additionally, there should be a relationship between the boundary Hilbert space defined by spin foam models \cite{sfbdy,sergei2} and the one which follows from a canonical quantisation, where one would have to deal with both types of constraint.

Therefore, we studied different formulations of the primary and secondary simplicity conditions in this work. We have seen that these different formulations interact differently with the degenerate sector that plays an important role in the quantum dynamics described by spin foam models \cite{nottingham}.

In particular, we derived gauge-variant simplicity constraints. In this form, the primary constraints \eqref{sec32d} are linear in the bi-vectors. It is in this guise that they recently led to proposals for new spin foam models \cite{epr, fk}. The secondary constraints \eqref{sec32e} involve the holonomies and, through an explicit construction of the spatial Levi-Civita connection, the bivectors in a complicated way. These constraints resemble discrete versions of the so-called reality conditions for (complex) Ashtekar variables, which ensure that the sum of self-dual and anti-self-dual connections is equal to (twice) the spatial Levi-Civita connection. 

A well-defined spatial Levi-Civita connection requires one to be on a proper geometric configuration, i.e. the gluing conditions \eqref{gluingb} must hold. For this reason, the gluing constraints actually {\it follow} from the secondary simplicity constraints (or, equivalently, from the edge simplicity constraints in the non-degenerate sector). The gluing conditions can be alternatively expressed as constraints \eqref{gluingc} ensuring that the extrinsic curvature, i.e. the 4d-dihedral angles are well-defined. This can be easily understood by realising that the self-dual or anti-self-dual connections are a linear combination of the spatial Levi-Civita connection and the extrinsic curvature. Hence, an ambiguity in the definition of the Levi-Civita connection  leads to an ambiguity in the definition of the  4d-dihedral angles and vice versa.

From just the primary simplicity constraints one might expect that the effect of the simplicity constraints is to reduce the $\SO(4)$ phase space to an $\SO(3)$ phase space. This is however not the case, since the gluing conditions (which are contained within the secondary simplicity constraints) reduce the $\SO(3)$ phase space even further.   If one uses the diagonal, cross and edge simplicity constraints \eqref{condis08},  it is essential that one requires non-degeneracy during this part of the reduction,. On the other hand, if one uses the secondary simplicity constraints in the form \eqref{sec32e} and in the form of the gluing constraints \eqref{gluingb}, one automatically removes the degenerate sector.  A further analysis of this issue would serve to benefit the assembly of methods to exclude non-degenerate configurations from spin foam models. 

In this work, we formulated different constraint sets and elucidated the relations between these various sets. We discussed the dimensionality of the reduced phase space for the case of the boundary of a 4-simplex. For more complicated triangulations, the issue is much more involved \cite{dr1} as one has to deal with non-local area constraints. Another issue is the derivation of the symplectic structure, i.e. the Dirac brackets for the reduced phase space. This will be the subject of further work \cite{dr3}, where we shall also comment on the role of the Immirzi parameter in the different stages of reduction.

%%%%%%%%%%%%%%
% Appendix
%%%%%%%%%%%%%%

\appendix

%%%%%%%%%%%%%%
% Conventions
%%%%%%%%%%%%%%

\section{Conventions}
\label{conv}

The discretisation of 4-dimensional $\SO(4)$ BF-theory relies heavily on the isomorphism between the space of bi-vectors (anti-symmetric second order tensors) and the algebra $\so(4)$.  In our conventions the commutation relations among the generators of the algebra take the form: 
\be
\{J^A,J^B\} = C^{ABC}\; J^C.
\ee
The indices are $A=a\bar{a}$ where $a,\bar{a} \in \{0,1,2,3\}$.  The structure constants are:
\be
\ba{rcl}
C^{ABC} &:=& \epsilon^{Ars} \; \epsilon^{Bst}\;\delta^{Ctr}\\[0.2cm]
&=&  \delta^{ab}\;\delta^{\bar{a}\bar{b}C} + \delta^{\bar{a}\bar{b}}\;\delta^{abC} - \delta^{a\bar{b}}\;\delta^{\bar{a}bC} - \delta^{\bar{a}b}\;\delta^{a\bar{b}C},
\ea
\ee
where the Kronecker delta and the summation convention on this space are given by:
\be
\ba{rcl}
\delta^{AB} & := & \delta^{ab}\delta^{\bar{a}\bar{b}} - \delta^{a\bar{b}}\delta^{\bar{a}b},\\[0.2cm]
T^{AB}\;U^{BC} & := & \frac{1}{2} T^{Ab\bar{b}} \;U^{b\bar{b}C} \quad\quad \textrm{so that} \quad\quad \epsilon^{AB}\;\epsilon^{BC} = \delta^{AC}.
\ea
\ee
We can use the Hodge $\star$ operator to split the algebra into self-dual and anti-self-dual sub-algebrae:
\be
J_{\pm}{}^A : = P_{\pm}{}^{AB} \; J^B, \quad\quad\textrm{where} \quad\quad P_{\pm}{}^{AB} := \frac{1}{2}(\delta^{AB} \pm\epsilon^{AB}).
\ee

It is straightforward to check that these projectors are orthonormal, that is $P_s^{AB}P_{s'}^{BC}=\delta_{ss'}P_s^{AC}$, furthermore $P_+{}^{AB}+P_-{}^{AB}=\delta^{AB}$. 
We may then proceed to generate all manner of projected quantities:
\begin{eqnarray}
X_\pm{}^A &=& P_\pm{}^{AA'}\; X^{A'}\;,\quad\quad\quad\quad M_\pm{}^A = P_\pm{}^{AA'}\;M^{A'}\;,\\
C_\pm{}^{ABC} &=& P_{\pm}{}^{AA'}\;C^{A'BC} = P_{\pm}{}^{BB'}\;C^{AB'C} = P_{\pm}{}^{CC'}\;C^{ABC'} = P_{\pm}{}^{AA'}\;P_{\pm}{}^{BB'}\;P_{\pm}{}^{CC'}\;C^{A'B'C'}\;.
\end{eqnarray}
The following identity for the $SO(4)$ structure constants:
\be\label{cons1}
C_\pm{}^{ABC}\; C_\pm{}^{A'B'C}=2^3 (P_\pm{}^{AA'}\; P_\pm{}^{BB'} - P_\pm{}^{AB'}\; P_\pm{}^{A'B} ).
\ee
 is in close analogy to the relation $\epsilon^{abc}\epsilon^{a'b'c}=(\delta^{aa'}\delta^{bb'}-\delta^{ab'}\delta^{a'b})$ for the $SO(3)$ structure constants.
 
 We can explicitly map from the self-dual and anti-self-dual sectors to corresponding $\SO(3)$ quantities in the vector representation. To this end, we use an auxiliary unit vector, which in this work is $e_t=(1,0,0,0)$. We then define:
\be
\ba{rcl}
x_{\pm}{}^r &=& 2 \;e_t{}^a\; P_{\pm{}}^{ar B} \; X^B,  \\[0.2cm]
m_{\pm}{}^{rs}  &=& 2\; e_t{}^a\; e_t{}^b\; P_{\pm}{}^{arB}\; P_{\pm}{}^{bsC} \;M^{BC} .  
\ea
\ee
where $r,s$ take values in $\{1,2,3\}$.

\section{Various representations of $\SO(4)$}
\label{reps}

Let us directly construct the relationship between various representations of $\SO(4)$ for future reference:
\begin{description}
\item[Vector:] The generators of the vector representation are:  $(J^A)^B = -\delta^{AB}$, and thus a generic transformation is given by:
\be\label{sec01}
M^{ab} = \exp( B\cdot J ){}^{ab}.
\ee
\item[Bi-vector:] One can then generate the bi-vector representation by taking the tensor product of two vector representations and antisymmetrising:
\be\label{sec02}
M^{AB} := 2\; M^{[a}{}_{[b}M^{\bar{a}]}{}_{\bar{b}]} = \exp(-B\cdot J)^{AB}, \quad\quad \textrm{where}\quad\quad (J^A)^{BC} = C^{ABC} .
\ee
\item[Quaternion:] The vector representation is isomorphic to the representation carried by the quaternions $\mathbb{H}$.  One maps a 4-vector $x = (x^0, x^1, x^2, x^3)$  to a pseudo-unitary matrix $x = x^{0}\sigma_0 + i x^r \sigma_r$.\footnote{
\be\label{sec03}
\sigma_0 := 
\left(\ba{cc}
1 & 0 \\
0 & 1
\ea\right), 
\quad\quad
\sigma_1 := 
\left(\ba{cc}
0 & 1 \\
1 & 0
\ea\right),
\quad\quad
\sigma_2 := 
\left(\ba{cc}
0 & -i \\
i & 0
\ea\right), 
\quad\quad
\sigma_3 := 
\left(\ba{cc}
1 & 0 \\
0 & -1
\ea\right). 
\ee
}
Then, the action of $\SO(4)$ on this space is:
\be\label{sec04}
x \rightarrow g_+ x g_-^{-1}, \quad\quad \textrm{where} \quad\quad 
\ba{rcl}
g_\pm &:=& \exp (-\frac{i}{2}b_{\pm}\cdot \sigma),\\[0.2cm]
b_\pm{}^r &:=& B_\pm{}^{0r}.
\ea
\ee
\item[Adjoint:] Furthermore, the bi-vector representation is isomorphic to the adjoint representation.   One maps a bi-vector to an element of $\so(4)$:  $X = X^{A}J^A$.  Then, $\SO(4)$ acts on its algebra by conjugation:
\be\label{sec05}
X \rightarrow G XG^{-1} = G_+X_+(G_+)^{-1}  + G_-X_-(G_-)^{-1}, \quad\quad \textrm{where} \quad\quad 
\ba{rcl}
G &:=& \exp (-\frac{1}{2}B\cdot J), \\[0.2cm] 
G_\pm &:=& \exp (-\frac{1}{2}B_{\pm}\cdot J_\pm).
\ea
\ee
\end{description}

\section{Continuum canonical formulation of Plebanski theory}\label{app3}

We describe some properties of an initial canonical analysis of the continuum theory. In \cite{henneaux}, where a comprehensive analysis of the Plebanski action is undertaken, they use a slightly different form of the simplicity constraints. It has the same class of non-degenerate solutions.  Here, upon performing a Legendre transform,\footnote{
In the canonical basis, the action takes the form:
\be\label{condis04}
\cS_{Pleb,\cM} = \int dx^4 \left[ \Pi_i{}^A\; \dot{w}_i{}^A + (\cP_{\gamma}\; X_{0i})^A\; F_i{}^A + w_0{}^A\; (D_i\; \Pi_i)^A + \tfrac{\gamma^2}{\gamma^2 - 1}\phi^{AB}\; X_{0i}{}^A\; (\cP_{-\gamma}\; \Pi_i)^B \right],
\ee
where $F_i{}^A := \frac{1}{2}\;\epsilon_{0ijk}\; F_{jk}{}^A$.
A priori, $X_{0i}$ is not a dynamical variable, along with $w_0$ and $\phi$ but to simplify the canonical analysis we give it a trivial dynamics by adding the terms: $P_i{}^A\; X_{0i}{}^A - \mu_i{}^A\; P_i{}^A$ to the action \eqref{condis04}.  This results in another primary constraint: 
\be\label{condis04aa}
P_i{}^A \approx 0
\ee
 in addition to \eqref{condis04a}.  Conservation of the Gau\ss~constraint does not lead to further constraints.  Conservation of \eqref{condis04aa} gives secondary constraints and equations which serve to determine the Lagrange multipliers $\phi^{AB}$ (meaning that simplicity constraints are second class).  Finally, conservation of the simplicity constraints leads to\eqref{condis04b} which in light of the extra constraint \eqref{condis04aa} has an additional term:
\be\label{condis04ab}
(\delta^{AC}\; \delta^{BC} - \tfrac{1}{6}\;\epsilon^{AB}\;\epsilon^{CD})\; \mu_i{}^C\;(\cP_\gamma\; \Pi_i)^D.
\ee
Thus, it leads to more secondary constraints and equations determining certain components of the $\mu_i{}^A$ Lagrange multpliers.  We shall not delve in the details here, but refer the reader to \cite{henneaux}.  We are interested just in the form of \eqref{condis04b}.
}
 one finds that the phase space is parameterised by:  $(w_i{}^A, \Pi_i{}^A := \frac{1}{2}\;\epsilon_{0ijk}\; \delta^{AB}\;(\cP_{\gamma}\; X_{jk})^B)$, where $\cP_{\gamma}{}^{AB} := \frac{1}{2}(\delta^{AB} +\frac{1}{\gamma}\epsilon^{AB})$.  The parameters $w_0{}^A$ and $\phi^{AB}$ are Lagrange multipliers imposing the primary constraints:
\be\label{condis04a}
\ba{rcl}
D_i\;\Pi_i{}_A & \approx &0,\\[0.2cm]
(\delta^{AC}\; \delta^{BC} - \frac{1}{6}\;\epsilon^{AB}\;\epsilon^{CD})\; X_{0i}{}^C\; (\cP_{-\gamma}\;\Pi_i)^D &\approx&0,
\ea
\ee
where $D_i$ is the covariant derivative with respect to the spatial components of the spin connection. The first constraint is the Gau\ss~constraint and generates $\SO(4)$ transformations, while the final constraint is the canonical form of the simplicity constraints.  One should then check the preservation of the constraint hypersurface under time evolution.  This yields secondary constraints.  We shall be interested in the secondary constraint arising from the preservations of the simplicity constraint above:
\be\label{condis04b}
(\delta^{AC}\; \delta^{BC} - \tfrac{1}{6}\;\epsilon^{AB}\;\epsilon^{CD})\; X_{0i}{}^C \; \Big[\epsilon_{0ijk}(\cP_{-\gamma}{}^2)^{DR}\; (D_j\; X_{0k})^R - C^{DRS}\; w_0{}^R\; (\cP_{-\gamma}\; \Pi_i)^S \Big],
\ee
where $C^{DRS}$ are the structure constants of $\so(4)$.  We note that this constraint is a functional of both the bi-vector and the connection. We shall see a reflection of this in the discrete theory.

\vspace{0.5cm}

~\\
{\bf \large Acknowledgements}
~\\
We thank Sergei Alexandrov, Daniele Oriti, Roberto Pereira and Simone Speziale for discussions. 
In particular we would like to thank Sergei Alexandrov for the suggestion that the gluing conditions should be part of the secondary simplicity constraints.

\end{document}